\documentclass[iop,numberedappendix]{emulateapj}
\usepackage{apjfonts} % Times fonts
\usepackage{graphicx}

\usepackage{amsmath}

\journalinfo{Astronomical~journal}
\def\lesssim{\mathrel{\hbox{\rlap{\hbox{\lower4pt\hbox{$\sim$}}}\hbox{$<$}}}}
\def\gtrsim{\mathrel{\hbox{\rlap{\hbox{\lower4pt\hbox{$\sim$}}}\hbox{$>$}}}}

\newcommand{\noprint}[1]{}

\shorttitle{MOJAVE XI: Spectral distributions}
\shortauthors{Hovatta et al.}

\begin{document}

\title{MOJAVE: Monitoring of Jets in Active Galactic Nuclei with VLBA
  Experiments. XI. Spectral distributions}


\author{Talvikki Hovatta}
\affil{Cahill Center for Astronomy \& Astrophysics, California
  Institute of Technology, 1200 E. California Blvd, Pasadena, CA
  91125, USA}
\affil{Aalto University Mets\"ahovi Radio Observatory, Mets\"ahovintie 114, 02540 Kylm\"al\"a, Finland}
\email{thovatta@caltech.edu}

\author{Margo F. Aller and Hugh D. Aller}
\affil{Department of Astronomy, University of Michigan, 830 Dennison Building, Ann Arbor, MI 48109-1042, USA}

\author{Eric Clausen-Brown}
\affil{Max-Planck-Institut f\"ur Radioastronomie, Auf dem
  H\"ugel 69, 53121 Bonn, Germany}

\author{Daniel C. Homan}
\affil{Department of Physics and Astronomy, Denison University, Granville, OH 43023, USA}

\author{Yuri Y. Kovalev}
\affil{Astro Space Center of Lebedev Physical Institute,
  Profsoyuznaya 84/32, 117997 Moscow, Russia}
\affil{Max-Planck-Institut f\"ur Radioastronomie, Auf dem
  H\"ugel 69, 53121 Bonn, Germany}

\author{Matthew L. Lister}
\affil{Department of Physics, Purdue University, 525 Northwestern Ave. West Lafayette, IN 47907, USA}

\author{Alexander B. Pushkarev}
\affil{Pulkovo Observatory, Pulkovskoe Chaussee 65/1, 196140
  St. Petersburg, Russia}
\affil{Crimean Astrophysical Observatory, 98409 Nauchny, Crimea, Ukraine}
\affil{Max-Planck-Institut f\"ur Radioastronomie, Auf dem
  H\"ugel 69, 53121 Bonn, Germany}

\author{Tuomas Savolainen}
\affil{Max-Planck-Institut f\"ur Radioastronomie, Auf dem
  H\"ugel 69, 53121 Bonn, Germany}


\begin{abstract}
We have obtained milliarcsecond-scale spectral index distributions for a sample of 190 extragalactic radio jets through the Monitoring of Jets in Active Galactic Nuclei with the VLBA Experiments (MOJAVE) project. The sources were observed in 2006 at 8.1, 8.4, 12.1, and 15.4\,GHz, and we have determined spectral index maps between 8.1 and 15.4\,GHz to study the four-frequency spectrum in individual jet features. We have performed detailed simulations to study the effects of image alignment and ({\it u,v})-plane coverage on the spectral index maps to verify our results. We use the spectral index maps to study the spectral index evolution along the jet and determine the spectral distributions in different locations of the jets. The core spectral indices are on average flat with mean value $+0.22\pm0.03$ for the sample, while the jet spectrum is in general steep with a mean index of $-1.04\pm0.03$. A simple power-law fit is often inadequate for the core regions, as expected if the cores are partially self-absorbed. The overall jet spectrum steepens at a rate of about $-0.001$ to $-0.004$ per deprojected parsec when moving further out from the core with flat spectrum radio quasars having significantly steeper spectra (mean $-1.09\pm0.04$) than the BL~Lac objects (mean $-0.80\pm0.05$). However, the spectrum in both types of objects flattens on average by $\sim 0.2$ at the locations of the jet components indicating particle acceleration or density enhancements along the jet. The mean spectral index at the component locations of $-0.81\pm0.02$ corresponds to a power-law index of $\sim 2.6$ for the electron energy distribution. We find a significant trend that jet components with linear polarization parallel to the jet (magnetic field perpendicular to the jet) have flatter spectra, as expected for transverse shocks. Compared to quasars, BL~Lacs have more jet components with perpendicular magnetic field alignment, which may explain their generally flatter spectra.
The overall steepening of the spectra with distance can be explained with radiative losses if the jets are collimating or with the evolution of the high-energy cutoff in the electron spectrum if the jets are conical. This interpretation is supported by a significant correlation with the age of the component and the spectral index, with older components having steeper spectra. 

\end{abstract}


\section{Introduction}
Relativistic jets of Active Galactic Nuclei (AGN) are thought to be formed via outflows generated when matter accretes 
around a supermassive black hole and gets expelled due to magnetic forces \citep[see][for a review]{meier01}.
Due to Doppler beaming the appearance of these sources in Very Long Baseline Interferometry (VLBI) observations is typically a one-sided 
jet \citep[e.g.,][]{kellermann04,lister09}.
Single-dish radio observations of these compact extragalactic objects often show 
remarkably flat continuum spectra dubbed as the ``cosmic conspiracy'' by \cite{cotton80}. Based on measurements 
over a wide range of frequencies, \cite{kellermann69} suggested this to be due to multiple homogeneous self-absorbed 
synchrotron components with a range of turnover frequencies, forming a flat overall 
spectrum. This hypothesis was confirmed by detailed VLBI observations of the compact flat-spectrum source 
0735+178 \citep{marscher77, cotton80}. Further interferometric observations showed that AGN viewed at small 
angles to the line of sight, generally called blazars, consist of a ``core'' with a flat or inverted spectrum and a jet with 
steep spectrum $\alpha \sim -0.7$ \citep[e.g.,][]{readhead79, marscher88,pushkarev12}. Throughout the paper we define the 
sign of the spectral index as $S \propto \nu^{+\alpha}$. 

Since the commissioning of the Very Long Baseline Array (VLBA) in 
  1993 it has become easier to conduct simultaneous 
multifrequency observations of the parsec-scale jets in AGN. \cite{osullivan09} studied the spectral distributions and magnetic fields in six blazars using VLBA observations at eight frequencies between 4.6 and 43\,GHz. They found the sources to be consistent with a Blandford - K\"onigl type conical jet \citep{blandford79} with an optically thick or self-absorbed core and an optically thin jet. Individual sources have been studied for their spectral distributions in great detail \citep[e.g.,][]{savolainen08,homan09,fromm13}.

Much work has also been done on the emission of radio galaxies 
\citep[e.g.,][]{walker00,vermeulen03,kadler04} and GPS (Gigahertz Peaked Spectrum) sources \citep[e.g.,][]{kameno00,marr01,tingay03,orienti08,orienti12}.
In these studies the spectra of the sources at various locations are extracted, and detailed modeling is done to investigate 
its nature. Assuming a power-law distribution of electrons $N(E) = N_0E^{-p}$, where $E$ is energy, the spectrum at high frequencies 
where the source is optically thin can be described with a power-law $I_\nu \propto \nu^\alpha$, where 
$\alpha = (1-p)/2$.
At low frequencies below the turnover, the spectrum does not depend on $p$.
The turnover can be either due to self-absorption of the synchrotron emitting electrons, in which case the low-frequency 
spectrum of a homogeneous optically thick source follows $I_\nu \propto \nu^{5/2}$, or due to free-free 
absorption in ionized gas surrounding the source, in which case the spectrum depends on the kinetic temperature and 
electron density of the ionized gas \citep{kellermann66}. \cite{walker00} used VLBA data at frequencies between 2.3 and 43\,GHz 
to study the inner regions of the radio galaxy 3C~84. They found a very steep spectral index of $+4$ below the spectral turnover in the northern 
counter-jet component and concluded that it must be due to free-free absorption by ionized gas associated with accretion disk. A similar conclusion 
had been drawn earlier by \cite{vermeulen94} based on non-simultaneous data at 8 and 22\,GHz. They argued that the surface 
brightness of the northern component was too low for the inverted spectrum to be due to synchrotron self-absorption. Another radio galaxy 
in which multifrequency VLBA observations have revealed free-free absorbed emission is NGC~1052 \citep{vermeulen03,kadler04}. It seems 
that free-free absorption due to ionized gas in the inner regions of AGN is fairly common and is best detected in nearby radio galaxies, 
where the angular resolution of the VLBA allows the study of the inner-most regions of the sources which can be obscured 
by the molecular torus, a source of ionized gas. Radio galaxies are also seen at a larger viewing angle, which allows the detection of the counter-jet and a better view to the nucleus than in blazars where the jet emission dominates.

Another interesting feature possibly observable in the spectra of synchrotron sources is the steepening of the spectrum 
due to synchrotron losses \citep{kardashev62}. This is called spectral aging and is typically expected at near-IR to optical wavelengths in 
newly emerged synchrotron components \citep[e.g.,][]{marscher85} but may be observable further down in the parsec-scale jets at radio frequencies. 
Most of the studies on spectral aging have been done in the kiloparsec scale lobes of radio galaxies where spectral steepening is 
typically observed. This is often interpreted to be due to synchrotron losses and spectral ages of the lobes can be calculated 
\citep[e.g.,][]{alexander87,carilli91,mack98,machalski09}. The ages derived this way are typically shorter than the age it has taken for the lobes to 
form and therefore re-acceleration is invoked (however, see a discussion on the caveats of the method in
\citealt{blundell01,rudnick02}). This is also a commonly used method to study the age of compact AGN such as 
GPS objects \citep[e.g.,][]{murgia03, nagai06} which support the interpretation that these sources are young objects 
\citep[see][for a review]{odea98}. 

Recently \cite{pushkarev12b} studied the spectra of parsec-scale jets in 319 compact AGN between 2 and 8\,GHz and found the 
mean spectral index in the jet to be $-0.68$. They also found that the spectral index steepens along the jet on average by 
$-0.06^{+0.07}_{-0.08}$~mas$^{-1}$. In this paper we will study the spectral evolution in the parsec-scale jets by examining the 
spectral index along the jet ridge lines and comparing the spectra of individual features to the time of the ejection of the component.

We present the spectral index maps of 190 sources studied also in \cite{hovatta12} and \cite{pushkarev12}. 
Details of our observations and methods used to derive the maps are given in Sect.~\ref{sect:data}. We study the spectral distributions in the parsec-scale core and jet components in Sect.~\ref{sect:dist}. Evolution of the spectrum as a function of distance and age is examined in Sect.~\ref{sect:aging}. We summarize with conclusions in Sect.~\ref{sect:conc}. 
Throughout the paper we use a cosmology where $H_0 = 71~\mathrm{km}\mathrm{s}^{-1}\mathrm{Mpc}^{-1}$,
$\Omega_M = 0.3$, and $\Omega_\Lambda = 0.7$ \citep[e.g.,][]{komatsu09}. 

\section{Observations and Methods}\label{sect:data}
MOJAVE (Monitoring of Jets in Active galactic nuclei with VLBA
Experiments) is an observing program to monitor the changes in a large sample of 
parsec-scale AGN jets in total intensity and polarization at 15.4\,GHz with the VLBA \citep{lister09}. 
In 2006 the monitoring was expanded to include multifrequency observations at 8.1, 8.4, 12.1, and 15.4\,GHz. 
Altogether 191 sources were observed (twenty of them twice) in batches distributed over 12 epochs. Our sample includes 133 flat-spectrum radio quasars (hereafter quasars), 
33 BL~Lac objects, 21 radio galaxies, and 4 optically un-identified objects. 
In addition to studying the spectral index 
distribution, these data were used to study Faraday rotation in these sources \citep{hovatta12} and 
the frequency-dependent shift of the optically thick base of the jet, the ``core-shift'' effect 
\citep{pushkarev12}.

The observations were made in dual polarization mode using 
frequencies centered at 8.104, 8.424, 12.119, and 15.369\,GHz. The bandwidths were 16 and 32\,MHz 
for the X and U-bands, respectively. The observations were recorded with a bit rate of 128 Mbits s$^{-1}$. In the 8\,GHz bands 
the observations consist of 2 sub-bands in both frequencies and 4 sub-bands in the 12 and 15\,GHz bands. All ten VLBA antennas were 
observing except at epoch 2006-Aug-09 when Pie Town was not included. The sources 
and their observing epochs are listed in Table~\ref{sptable} where column (1) gives the B-1950 name of the source, column (2) the alternative name, column (3) the redshift of the object, column (4) the optical classification, column (5) the maximum apparent speed for the source, column (6) the date of our observation, column (7) the spectral index in the core component, column (8) the spectral index in the jet, and column (9) the alignment flag described in Sect.~\ref{sect:align}. 

\begin{table*}\scriptsize
\begin{center}
\caption{\label{sptable}Sources and their spectral index properties}  
\begin{tabular}{lccccrrrc} 
\tableline\tableline
IAU name & Other name & z & Opt. Cl. & $\beta_{app}$ &  
Epoch & core $\alpha$  & jet $\alpha$ & alignment flag \\  
& & & & (c) & & & & \\ 
(1) & (2) & (3) & (4) &  
(5) & (6) & (7) & (8) & (9) \\ 
\tableline 
0003$-$066 & NRAO 005 & 0.3467 & B & 8.4 & 2006 JUL 07 & $-0.04$ & $-1.01$ &  \\ 
0003+380 & S4 0003+38 & 0.229 & Q & 4.6 & 2006 MAR 09 & $-0.03$ & $-0.80$ &  \\ 
0003+380 & S4 0003+38 & 0.229 & Q & 4.6 & 2006 DEC 01 & $-0.24$ & $-1.20$ &  \\ 
0007+106 & III Zw 2 & 0.0893 & G & 1.2 & 2006 JUN 15 & $-0.47$ & \nodata & Y \\ 
0010+405 & 4C +40.01 & 0.256 & Q & 6.9 & 2006 APR 05 & $0.25$ & \nodata & Y \\ 
0010+405 & 4C +40.01 & 0.256 & Q & 6.9 & 2006 DEC 01 & $0.16$ & \nodata & Y \\ 
0016+731 & S5 0016+73 & 1.781 & Q & 8.2 & 2006 AUG 09 & $0.55$ & \nodata & Y \\ 
\tableline
\end{tabular}
\tablecomments{Columns are as follows: (1) IAU Name (B1950); (2) Other name; (3) redshift; (4) optical classification where Q = quasar, B = BL Lac object, G = active galaxy, and U = unidentified; (5) Apparent speed used in viewing angle calculation of Fig. \ref{fig:age}, taken from \cite{lister13}; (6) epoch of the spectral index observation; (7) Spectral index in the core component; (8) Median spectral index over the jet ridge line; (9) Y if 2D cross-correlation not possible for image alignment (see sect.~\ref{sect:align} for details);\\
(This table is available in its entirety in machine-readable and Virtual Observatory (VO) forms in the online journal. A portion is shown here for guidance regarding
its form and content.)}
\end{center}
\end{table*}

\subsection{Data reduction}
The initial data reduction and calibration were performed following the standard procedures described 
in the AIPS cookbook\footnote{http://www.aips.nrao.edu}. All the frequency bands were 
treated separately throughout the data reduction process. The imaging and self-calibration
were done in a largely automated way using the Difmap package \citep{shepherd97}. 
One source, 0108+388, had no single bright component to be used in self-calibration and was dropped from the analysis. 
For more details see \cite{lister09} for the MOJAVE data reduction and imaging process.

The absolute flux density calibration accuracy was checked against the single-dish monitoring data at 8 and 14.5\,GHz from the University of Michigan Radio Astronomy Observatory (UMRAO). At 8 and 15\,GHz we expect the absolute calibration uncertainty to be 5\% \citep{lister05}. At 12\,GHz we did not have independent single-dish observations of the sources and originally the same calibration factors as for 15\,GHz were used. The calibration of the 12\,GHz data was examined using the total integrated flux density of the sources.  
We fit a power-law model to the data with the 12\,GHz data excluded. A scaling factor was then defined from the fits, by interpolating the expected flux density and calculating the ratio to the original data. A median value was determined for each epoch and applied to the 12\,GHz data.
The scatter in the scaling factors for all epochs was about 7.5\%, and we adopt this as our absolute calibration uncertainty for the 12\,GHz band.

All the maps were modeled with circular or elliptical Gaussian components using 
the standard procedure in the Difmap package. The 15\,GHz maps were previously modeled 
as a part of the MOJAVE survey \citep[][hereafter Paper X]{lister13}. Since one of our goals was 
to use the optically thin components in the jets to align our images, 
we used these 15\,GHz models as a starting point for the other bands 
and modified the fit if needed. 

The ({\it u,v})-plane coverage differs between the 8, 12, and 15\,GHz bands, with 12 and 15\,GHz band data having longer 
baselines resulting in better resolution. On the other hand, the 8\,GHz data has shorter baselines 
making it more sensitive to extended emission. This may result in steepening of the spectral 
indices in the extended jet regions where the 12 and 15\,GHz observations are not as sensitive.
Therefore, in order to have comparable ({\it u,v}) 
coverage in all the bands, we deleted the long baselines from the 15 and 
12\,GHz maps and short baselines from the 8\,GHz maps. The resulting typical  ({\it u,v}) 
range in our data is 7.3 - 231 M$\lambda$.
Additionally, we restored all the maps to the 
beam size of our lowest frequency (8.1\,GHz).
All these steps were carried out in Difmap. We further study the effect of the ({\it u,v})-plane coverage on our spectral index maps by detailed simulations, described in Appendix~\ref{app:steep}.

\subsection{Image alignment}\label{sect:align}
During the self-calibration process the absolute coordinate position of the source is lost and 
the brightest feature of the image is shifted to the phase center of the map. This may not be the 
same position on the sky for different frequency bands, and therefore an extra step is needed to align the images.
This can be done using bright components in the optically thin part of the jet, 
whose position should not depend on the observing frequency \citep[e.g.,][]{marr01,kovalev08, sokolovsky11,fromm13}. This approach works well for knotty 
jets but is unreliable or impossible to use for faint or smooth jets. A solution 
is to use a 2D cross-correlation algorithm to look for the best alignment based on correlation of 
the optically thin parts of the jets at different bands \citep{walker00,croke08,fromm13}. 

In order to diminish the effect of errors in the alignment, we used both optically thin bright components and 
2D cross-correlation to align the images whenever possible. All the shifts were 
verified by examining the spectral index maps before and after the alignment. In 46 cases we were not able to 
find a reliable alignment due to the compactness of the source or the faintness of a featureless jet (note that the number is larger than reported in \citet{hovatta12} because in that paper we considered only sources with detected polarized emission). 
In these cases we aligned the images based on the fitted core component position 
at each band or on the phase center, whichever resulted in a smoother map. 
These sources are marked in column (9) of Table~\ref{sptable} and their spectral index values should in general be considered less reliable than the sources with proper 2D cross-correlation alignment. 

The image alignment was originally done by \cite{hovatta12}, where we tested the effect of wrong alignments on 
the spectral index and rotation measure maps. This was done by introducing fake shifts between images at 
different bands and visually examining the resulting spectral index and rotation measure maps. We noticed that the 
effect of incorrect alignment is stronger in spectral index maps and therefore extra care needs to be taken 
when aligning the images (see also Fig. 7 in \cite*{kovalev08} for an example). In order to test the effect further we performed simulations and tests which are described in Appendix~\ref{app:align}. Based on the tests we conclude that the errors introduced by the 2D cross-correlation method dominate over the thermal noise up to a distance of 3\,mas from the phase center after which they approach the median value of 0.05. In the majority of the sources the errors are small but one should be cautious when making conclusions on the core region spectral index distribution in individual sources.

\subsection{Spectral index maps}\label{sect:maps}
\begin{figure*}[ht!]
%\epsscale{.80}
\begin{center}
\includegraphics[angle=-90,scale=0.7]{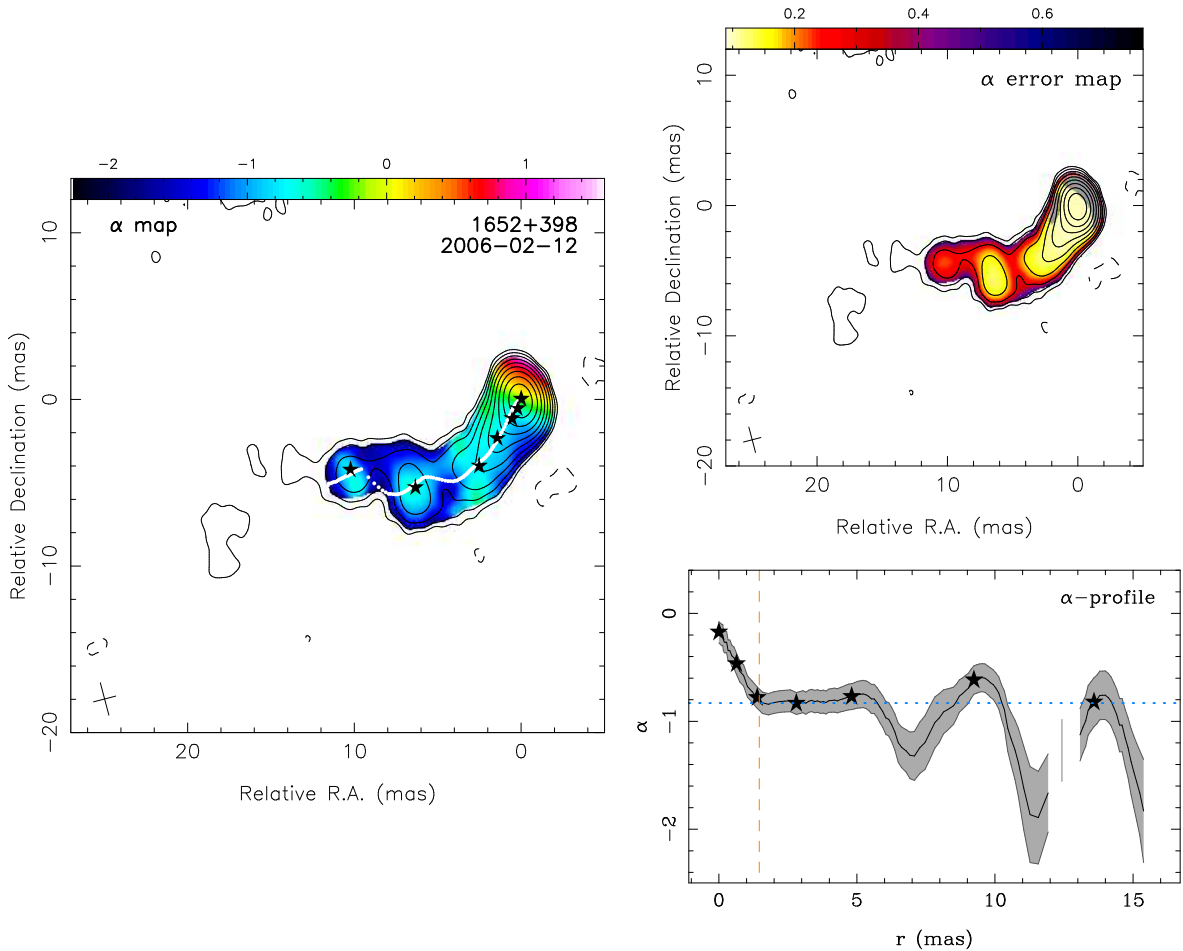}
\caption{Spectral index maps of all sources. The spectral index is shown in color on the left, overlaid on the 15\,GHz total intensity contours. White dots on the map show the locations of the ridge line points and black stars indicate the centroid positions of the fitted Gaussian components. The spectral index error map is shown in the top right panel. The error is shown in black (gray in the online journal) if the spectral index fit has $\chi^2 > 6$. The bottom right panel shows the spectral index values along the ridge line if the ridge line extends at least one beam size away from the convolved core. The dotted horizontal line (blue in the online journal) indicates the median jet spectral index and the dashed vertical line (orange in the online journal) the edge of the convolved core. The black stars show locations of the fitted Gaussian components if they are less than 0.3\,mas from the ridge line. The gray area indicates 1$\sigma$ errors on the spectral index. (The complete figure set (210 images) and color version of this figure are available in the online journal.)} \label{spmap}
\end{center}
\end{figure*}

The spectral index maps between 8.1 and 15.4\,GHz for all the sources are presented in Figs.~\ref{spmap}.1-~\ref{spmap}.210. The spectral index in each pixel was calculated 
by fitting a power-law to the total intensity data using the four frequency bands. 
We blanked pixels where the total intensity level was less than 3$\sigma_\nu$ at the given band, where 
$\sigma_\nu$ is defined as
\begin{equation}\label{eq:err}
\sigma_\nu = \sqrt{\sigma_\mathrm{rms}^2 + (1.5
  \sigma_\mathrm{rms})^2} \approx 1.8\sigma_\mathrm{rms},
\end{equation}
where $\sigma_\mathrm{rms}$ is the thermal noise of the image taken at a location 1 arcsec away from the center of the 
map. The second term under the squaroot sign in Eq.~\ref{eq:err} accounts for uncertainties due to the CLEAN procedure \citep[see][Appendix B for details]{hovatta12}. In calculating the fits we also added in quadrature an absolute calibration uncertainty of 5\% for the 8 and 15\,GHz bands and 7.5\% for the 12\,GHz band. The uncertainties of the spectral index are calculated from the variance-covariance matrix of the least-squares fit. Our fits have two degrees of freedom and we use a 95\% limit of $\chi^2 < 5.99$ from the $\chi^2$ distribution to determine if the fit is good. Regions with $\chi^2 > 6$ are shown in gray on the error maps in Figs.~\ref{spmap}.1-~\ref{spmap}.210.

There are several sources that show very large areas of bad $\chi^2$ fits even in the jets, where we expect a power law to be a good representation of the spectral index. The two main causes for this are the uncertainty of the 12\,GHz scaling and the uncertainty in the image alignment. By examining a few such cases individually, we conclude that the effect on the spectral index value is small.

The extraction of spectral information in the sources can be done in two ways by using either 1) the visibility data and 
fitting the source structure with several Gaussian features or 2) the image-plane data and extracting information from 
the convolved maps. In order to use method 1) the features at different frequencies have to be cross-identified. 
One way to do this is to use a higher frequency fit and transfer that to lower frequencies by keeping the 
positions and sizes of the components fixed and letting only the flux density to vary \citep{homan02, savolainen08, homan09}. 
This allows one to extract the spectrum of individual jet features, but it is a very time consuming method because of 
the complexity of the jets and the intrinsic differences between the frequency bands. \cite{homan09} solved this by 
modeling the sources with multiple different fits to obtain an average fit between the frequencies. This 
approach is good for studies of individual sources but is not feasible for samples as large as ours. Therefore we have 
used method 2) and extracted the information from convolved maps. 

\section{Spectral distributions}\label{sect:dist}
The spectral distribution along the source can be studied in several ways. As a first approach we extracted the spectral index at the fitted 15\,GHz Gaussian component locations. We calculated the average spectral index over a $3\times3$ box of pixels centered around the component position to avoid making conclusions based on single pixel values. With a pixel size of 0.1~mas, this corresponds to 10-30\% of the restoring beam width, depending on the declination of the source. The component locations and their spectral index value are given in Table~\ref{comptable} and shown over the spectral index maps in Figs.~\ref{spmap}.1.-\ref{spmap}.210. In Table~\ref{comptable}, column (1) gives the source name, column (2) the I.D. number of the component, column (3) the date of our observation, column (4) the distance of the component from the phase center of the map, column (5) the position angle of the component from the phase center, column (6) the spectral index of the component, column (7) the age of the component (see Sect.~\ref{sect:aging}), column (8) the fractional polarization, column (9) the electric vector position angle, and column (10) indicates if the component is more than a beam size away from the core.
The location of the core in the source 1404+286 is uncertain \citep{wu13}, and we exclude the source from all the further analysis that requires the core position to be known. In the source 0238$-$084 (NGC~1052) the core cannot be detected due to free-free absorption, and a virtual core is used in determining the kinematics (Paper X), so only jet components are used in our analysis for this source.

\begin{table*}\scriptsize
\begin{center}
\caption{\label{comptable}Fitted Gaussian components at 15.4 GHz and their spectral index values}  
\begin{tabular}{lccccccccc} 
\tableline\tableline
Source &I.D. & Epoch & r &  
P.A. & $\alpha$ & Age & Frac. pol & EVPA & jet flag\\  
 &  &  & (mas) &  
(deg) &  & (years) & (\%) & (deg.) &  \\ 
(1) & (2) & (3) & (4) &  
(5) & (6) & (7) & (8) & (9) & (10) \\
\tableline
0003$-$066 & 0 & 2006 JUL 07 & 0.7 & $-168.6$ & $-0.04\pm0.1$ & \nodata & \nodata & \nodata & \nodata \\ 
0003$-$066 & 1 & 2006 JUL 07 & 0.7 & $-71.6$ & $-0.18\pm0.1$ & \nodata & \nodata & \nodata & \nodata \\ 
0003$-$066 & 4 & 2006 JUL 07 & 6.9 & $-81.1$ & $-0.94\pm0.1$ & \nodata & $8.6\pm1.1$ & $-79.0\pm5.2$ & Y \\ 
0003$-$066 & 5 & 2006 JUL 07 & 0.1 & $2.9$ & $-0.09\pm0.1$ & \nodata & \nodata & \nodata & \nodata \\ 
0003$-$066 & 6 & 2006 JUL 07 & 1.3 & $-102.7$ & $-0.45\pm0.1$ & \nodata & \nodata & \nodata & \nodata \\ 
0003+380 & 0 & 2006 MAR 09 & 0.0 & $-69.3$ & $-0.03\pm0.1$ & \nodata & \nodata & \nodata & \nodata \\ 
0003+380 & 1 & 2006 MAR 09 & 3.9 & $121.9$ & $-0.54\pm0.3$ & \nodata & \nodata & \nodata & Y \\ 
\tableline
\end{tabular}
\tablecomments{Columns are as follows: (1) IAU Name (B1950.0); (2) I.D. of the component (0 = core); (3) Epoch; (4) Component distance from the phase center of the I map; (5) Position angle of the component from the phase center; (6) Component spectral index; (7) Component age calculated as the difference between the observing epoch and the ejection epoch from Paper X; (8) Component fractional polarization; (9) Faraday-corrected electic vector position angle of the component; (10) Y if the component is at least one beam size away from the core.\\
(This table is available in its entirety in machine-readable and Virtual Observatory (VO) forms in the online journal. A portion is shown here for guidance regarding
its form and content.)}
\end{center}
\end{table*}

Another approach we have used is to calculate the jet ridge line and extract the 
spectral index values along the ridge line. First, using the 15\,GHz data, the total intensity ridge line
was constructed with nearly equally spaced points separated by a
distance of the pixel size. The procedure starts from the core component position and continues until the peak of a Gaussian
curve fitted to the transverse total intensity jet profile becomes less than 4 rms noise level of the image. Applying the obtained ridge
line to the spectral index distribution map, we extracted the values of $\alpha$ along the jet. After that, we masked out the core region
within $r_\mathrm{core} < (b^2_\varphi + d^2_\varphi)^{0.5}$, where $b_\varphi$ and $d_\varphi$ are the FWHM of the restoring beam and fitted
core size, respectively, calculated along a position angle $\varphi$ of the inner jet. We show the ridge line in Figs.~\ref{spmap}.1.-\ref{spmap}.210 if the jet ridge line extends at least one beam size away from the convolved core region. For these sources, the median jet spectral index derived from the ridge line, excluding the convolved core region, is tabulated in Table~\ref{sptable}. In a few cases we were not able to determine a robust ridge line due to uncertainty in the core location (1404+286), long complicated jet (0429+415) or a complex two-sided jet (0238$-$084, 1509+054, 1957+405).

\subsection{Spectral index distribution in the cores}
\begin{figure}[ht!]
%\epsscale{.80}
\begin{center}
\includegraphics[angle=-90,scale=0.6]{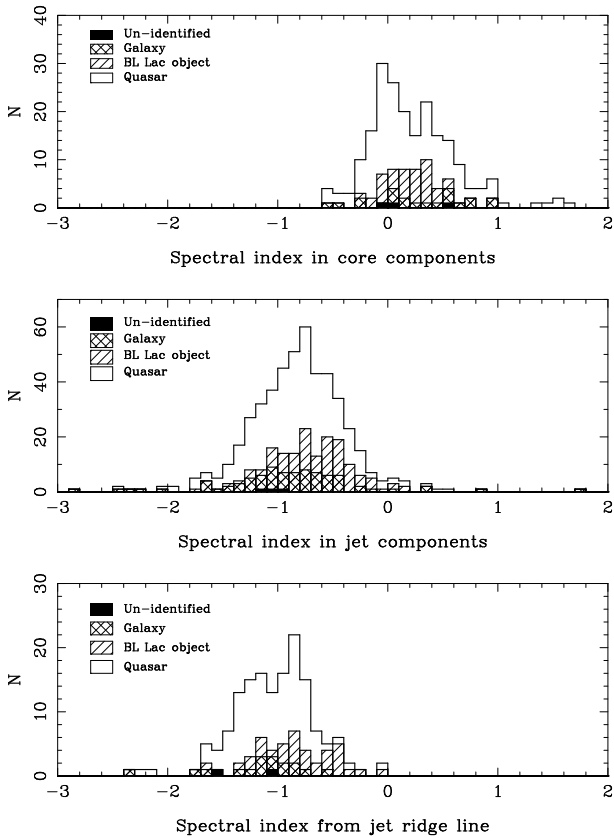}
\caption{Distribution of spectral index in the core components (top), jet components (middle), and jet ridge lines (bottom). The spectral index values are derived from the component and ridge line locations in the image plane. The x-axis of the middle panel excludes a jet component of the source 0238$-$084 on 2006-Dec-01 with a spectral index of 3.6.\label{fig:corejet}}
\end{center}
\end{figure}

The cores of AGN at cm wavelengths are considered to be the $\tau = 1$
surface of the jet where the emission transitions from optically thick
to optically thin \citep[e.g.,][]{blandford79,marscher80}.  If the energy density of the
synchrotron emitting particles is a constant fraction of the magnetic
energy density in this region, then the radio core is predicted to
have a flat spectral index, $\alpha = 0$, though this result depends
on how the magnetic field energy density decays with distance
\citep{blandford79,konigl81}.  On the other hand, if a constant particle-to-magnetic
energy density ratio cannot be realized in the radio core due to a
lack of particle reacceleration, then particle energy losses due to
adiabatic expansion predict an inverted spectral index of $\alpha \sim
0.5$ to 1 \citep{marscher80}, depending on various model assumptions.  As
seen in top panel of Fig. \ref{fig:corejet}, the distribution peaks at $\alpha \sim 0$, lending support to the constant particle-to-magnetic energy density model.  However, the distribution of $\alpha$ includes a large number of sources with inverted spectral indices, suggesting the adiabatic losses model may also be realized in some jets.
According to the non-parametric Kolmogorov-Smirnov (K-S) test, the probability that the 
quasars (mean $0.22 \pm 0.04$) and BL~Lacs (mean $0.19 \pm 0.03$) come from the same population is $p=0.034$, indicating that the classes differ at a $2\sigma$ level. 

It is worth pointing out that our cores are in some cases resolved, revealing a more complicated structure that cannot be described by a single core spectral index as is done in the jet models described above. Many of the spectral index maps show that the spectral index in the core changes smoothly from more inverted values to flat. The smoothness of this transition is due to the convolution with the finite beam, but as shown in the simulations in Appendix~\ref{app:core}, the change in the spectral index values is real and intrinsic to the source. This means that we can typically detect the transition from inverted synchrotron self-absorbed spectra to flat and optically thin. Unfortunately the image alignment errors can be large, especially in the core region, which prevents us from modeling the transition in more detail. We note that in many sources a simple power law is not a good spectral fit to the core regions, further implying that they are at least partially self-absorbed.

\begin{figure}[ht!]
%\epsscale{.80}
\begin{center}
\includegraphics[scale=0.45]{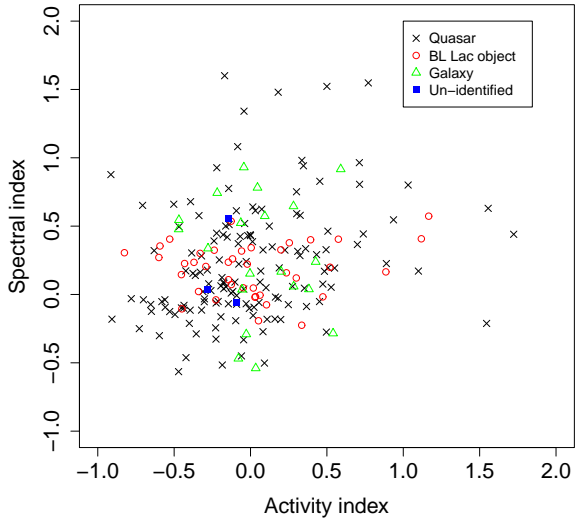}
\caption{Spectral index of the core components against the activity index. Black crosses are quasars, open circles (red in the online journal) BL Lac objects, open triangles (green in the online journal) radio galaxies, and squares (blue in the online journal) objects with unknown optical classification.
(A color version of this figure is available in the online journal.)\label{fig:activity}}
\end{center}
\end{figure}

As noted above, several core components show highly inverted
  spectra of $\alpha > 0.5$. This could also be a sign of new optically thick components within the convolved core. In order to study if the more inverted core spectra are due to a higher activity state of the source, we take advantage of the MOJAVE monitoring data and calculate the activity index \citep[e.g.,][]{kovalev09}, defined as
\begin{equation}
V = \frac{S - <S>}{<S>},
\end{equation}
where $S$ is the flux density of the core at 15\,GHz at the epoch of the spectral index measurement and $<S>$ is the mean flux density of the core. For each source, in the calculation of $<S>$, we used all available epochs between 1994 and 2013 in the MOJAVE data archive\footnote{http://www.physics.purdue.edu/astro/MOJAVE/allsources.html} (Paper X).
The correlation between spectral index and activity index is shown in Fig.~\ref{fig:activity}. According to the non-parametric Kendall's tau test, there is a significant correlation with $\tau_K=0.15$, $p=0.001$ between the two parameters. This agrees with multi-epoch observations of the quasar 2230+114 by \cite{fromm13} who found the core spectral index to be more inverted when a large total intensity flare was observed in the source. The correlation is driven by the quasars ($\tau_K=0.22$, $p=0.0001$) while there is no correlation for the BL~Lac objects alone ($\tau_K=-0.02$, $p=0.90$). 

\begin{figure}[ht!]
%\epsscale{.80}
\begin{center}
\includegraphics[scale=0.45]{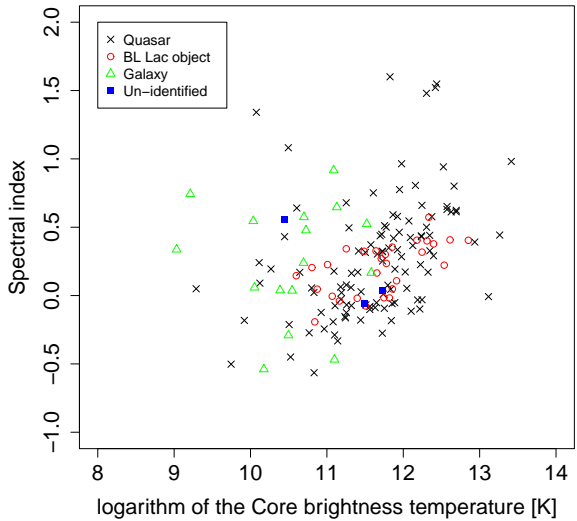}
\caption{Spectral index of the core components against the core brightness temperature. Black crosses are quasars, open circles (red in the online journal) BL Lac objects, open triangles (green in the online journal) radio galaxies and squares (blue in the online journal) objects with unknown optical classification.
(A color version of this figure is available in the online journal.)\label{fig:Tb}}
\end{center}
\end{figure}

Another measure of the activity state of the object is the brightness temperature. Fig.~\ref{fig:Tb} shows the spectral index of the core against its brightness temperature calculated using,
\begin{equation}\label{eq:Tb}
T_\mathrm{b,VLBI} = 1.22\times10^{12}\frac{S(1+z)}{\theta_\mathrm{min}\theta_\mathrm{maj}\nu^2} \mathrm{K},
\end{equation}
where $S$ is the flux density of the component in Jy, $z$ is the redshift, $\nu$ is the frequency in GHz (in our analysis 15.4) and $\theta_\mathrm{min}$ and $\theta_\mathrm{maj}$ are
the minor and major axis size of the component in mas. The correlation is highly significant ($\tau_K=0.27$, $p=4.8\times10^{-7}$) for all the sources and for quasars ($\tau_K=0.36$, $p=1.7\times10^{-8}$) and BL~Lacs ($\tau_K=0.42$, $p=0.001$) separately. 

\begin{figure}[ht!]
%\epsscale{.80}
\begin{center}
\includegraphics[scale=0.7]{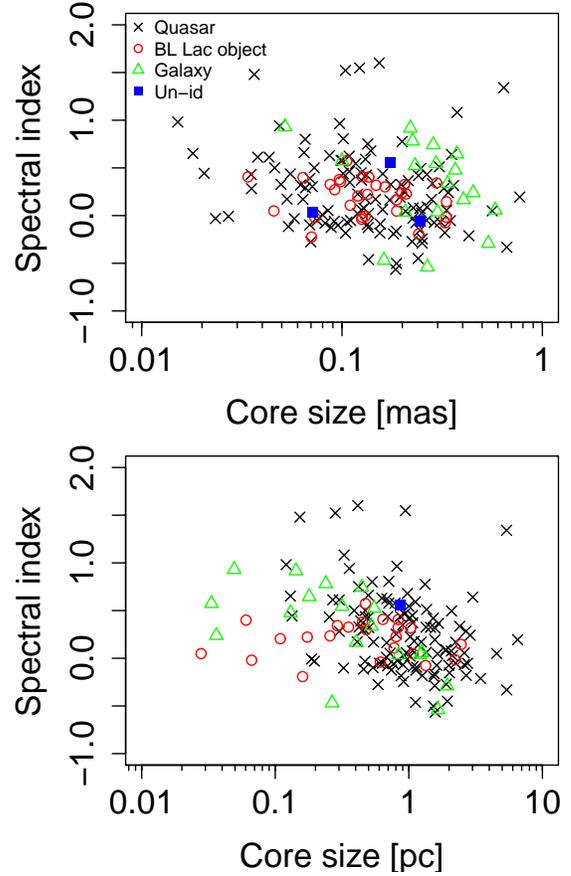}
\caption{Spectral index of the core components against the fitted core
  size in mas (top) and pc (bottom). Black crosses are quasars, open circles (red in the online journal) BL Lac objects, open triangles (green in the online journal) radio galaxies and squares (blue in the online journal) objects with unknown optical classification.
(A color version of this figure is available in the online journal.)\label{fig:size}}
\end{center}
\end{figure}

This correlation could be due to less contamination from the optically
thin jet in those core components that have higher brightness
temperatures. Indeed, in quasars, we see a significant negative
correlation ($\tau_K=-0.17$, $p=0.007$ ) between the core spectral
index and the angular size of the core, as shown in
Fig.~\ref{fig:size} top panel. This may indicate that the core size in some sources is overestimated, which if true, causes the apparent brightness temperatures to be underestimated. The effect is even more pronounced ($\tau_K=-0.23$, $p=9.3\times10^{-7}$) if
we take the redshift of the sources into account and calculate the
size of the core in parsecs, as shown in Fig. ~\ref{fig:size} bottom
panel. This shows that in sources at higher redshifts where the
effective linear resolution is poorer, there is more likely to be contamination
from the optically thin jet.

\subsection{Spectral index in the jets}
The distribution of the jet component spectral indices is shown in Fig.~\ref{fig:corejet} 
middle panel. We include here only components that are more than one beam width away from the core component to account for the contribution of the core in the convolved images (see Appendix~\ref{app:core} for justification of this choice). The mean value of all the sources ($-0.81 \pm 0.02$) is 
close to the value $-0.68$ obtained between 2 and 8\,GHz for a 
sample of 319 compact AGN \citep{pushkarev12b}. 
However, according to a K-S test the probability that quasars (mean $-0.85 \pm 0.02$) 
and BL~Lacs (mean $-0.64 \pm 0.03$) come from the same population is $p<1.3\times10^{-5}$. A Wilcox generalized rank test gives a probability of $p<2\times10^{-7}$
that quasars have similar jet component spectra as BL~Lacs. 

One reason for the steeper spectra in quasars could be the different redshift distributions of BL~Lacs and quasars. We test this by comparing the jet spectral indices in sources with $0 < z < 0.5$, the range of redshifts for most BL~Lacs in our sample. The mean spectral index for quasars in this range is $-0.84\pm0.06$, which differs significantly (K-S test $p=0.0047$, Wilcox test $p=0.0006$) from the BL~Lacs with a mean index of $-0.60\pm0.04$. This indicates that the difference in quasars and BL~Lacs is not simply due to a redshift effect.

In the bottom panel of Fig. ~\ref{fig:corejet} we show the median jet spectral index values determined from the ridge lines. The mean value is significantly (K-S test p-value $< 5\times10^{-9}$) steeper (mean $-1.04\pm0.03$) than the jet component value. The difference between quasars (mean $-1.09 \pm 0.04$) and BL~Lacs (mean $-0.79 \pm 0.05$) is still significant according to both K-S ($p=0.0026$) and Wilcox ($p=3.3\times10^{-5}$) tests. 

The difference between the component and ridge line spectral indices
indicates that the spectral index flattens at the component
locations. This can be also seen in Fig.~\ref{spmap} when comparing
the ridge line spectral index values to the component locations shown
by stars. To study this further we have calculated the difference
between the two values for each jet component, which describes the
amount of flattening at the component locations. The mean value for
all the components is $0.20\pm0.02$. The difference between quasars
(mean $0.21\pm0.02$) and BL~Lacs (mean $0.15\pm0.03$) is not
significant ($p=0.15$) according to a K-S test. The most simple
explanation for the flattening is that the components represent
physical structures in the jets where on-going particle acceleration
flattens the spectrum. This would be expected, for example, for shocks in the jets, based on hydrodynamical simulations of relativistic jets \citep{mimica09}. In Fig.~\ref{spmap}.200 (available in the online edition of the journal), the ridge line plot of the source 2230+114 shows such a bump about 5\,mas away from the core. \cite{fromm13} studied the jet of 2230+114 over a wide frequency range and suggest that the feature is a recollimation shock in the jet and interpret it in the framework of the \cite{mimica09} simulations. 

The observed spectral index of $\alpha=-0.8$ at the component locations corresponds to a power-law index of $p=2.6$ for the electron distribution. This is softer than the ``canonical'' index 2.2, but consistent with theoretical expectations for the first-order Fermi acceleration in relativistic shocks with finite thickness \citep[e.g.,][]{kirk87,virtanen05}.

In order to further test if shocks are responsible for flatter spectra, we study the relation between spectral index and polarization in the jet components that are at least one beam width away from the core. We use the position angle (PA) of the component relative to the core as a measure of the local jet direction. We then calculate the absolute difference between the PA and the Faraday rotation-corrected electric vector position angle (EVPA) of the component. The Faraday rotation measure values are taken from \cite{hovatta12} and we only include components for which an estimate of the Faraday rotation was available. We tabulate the corrected EVPA values and fractional polarization for the components in this plot in Table~\ref{comptable}.

\begin{figure}[ht!]
%\epsscale{.80}
\begin{center}
\includegraphics[scale=0.6]{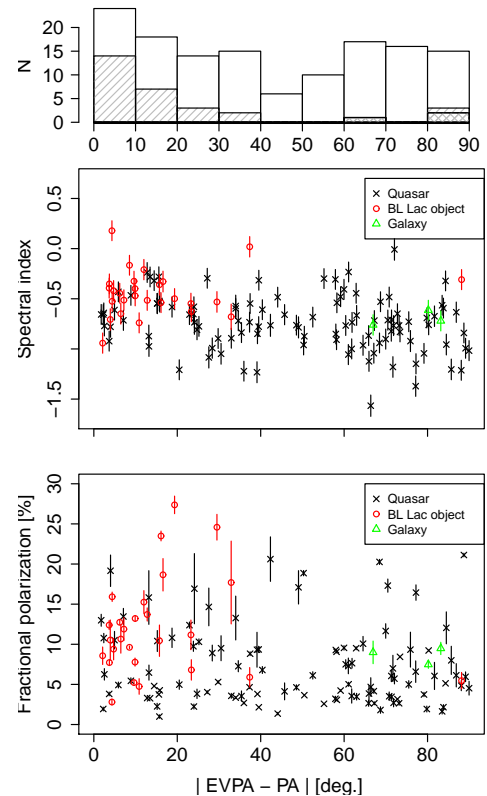}
\caption{Top: Distribution of the absolute difference between the position angle and the Faraday rotation-corrected EVPA of the component. Hatched regions are BL~Lac objects, cross-hatched regions are galaxies and white regions quasars. Middle: Spectral index of the component against the EVPA - PA difference. Bottom: Fractional polarization of the component against the EVPA - PA difference. Black crosses are quasars, open circles (red in the online journal) are BL~Lac objects and triangles (green in the online journal) are galaxies.
(A color version of this figure is available in the online journal.)\label{fig:pol}}
\end{center}
\end{figure}

In optically thin jets, we expect the magnetic field to be
  perpendicular to the EVPA, and therefore this measure will tell us
  the direction of the magnetic field with respect to the jet
  direction. In Fig.~\ref{fig:pol} top panel we show the distribution
of the EVPA - PA difference for the different subclasses. All except
one BL~Lac object (0851+202), are seen at angles of less than 40
degrees while the quasars occupy the entire range. This indicates that
in BL~Lac objects the magnetic field is predominantly perpendicular to
the jet direction, as expected for shocks with fronts oriented transverse to the jet axis or a helical magnetic field \citep[e.g.,][]{gabuzda94}. This distribution is very similar to what was seen in the first MOJAVE epoch \citep{lister05} for the jets of quasars and BL~Lacs. 
We note that some events in the single-dish light curves of 0851+202 require an oblique shock orientation when modeled in detail \citep{aller13}, which may indicate that the division between BL~Lac objects and quasars does not always hold for individual objects.

In the middle panel of Fig.~\ref{fig:pol} we plot the spectral index
against the EVPA - PA difference. The BL~Lacs have significantly
flatter spectra than the quasars but there is also a negative trend in
the quasars alone, with components with smaller difference having flatter spectra ($\tau_K=-0.17$, $p=0.009$). Another test is to compare the spectral indices at angles $<20^\circ$ (mean index $-0.59\pm0.05$) and at angles $>70^\circ$ (mean index $-0.81\pm0.05$), which according to a K-S test are coming from the same population with a probability $p=0.008$. This implies that components where the magnetic field is more perpendicular to the jet direction, a signature of shocks, have flatter spectra supporting our hypothesis that the flatter spectra are due to shocks in the jets.

As suggested by \cite{aller99} and confirmed with hydrodynamic
simulations by \cite{hughes05}, this may also imply that shocks form
more readily in BL~Lacs jets than quasars, explaining their generally
flatter spectra. In Fig.~\ref{fig:pol} bottom panel we plot the
fractional polarization of the component against the EVPA - PA
difference. The fractional polarization of BL~Lac objects (mean
$11.0\pm1.1$) is significantly higher than in the quasars (mean
$6.7\pm0.4$) according to a Wilcox test ($p=5.6\times10^{-5}$), and
according to a K-S test they come from the same population with a
probability of $p=0.0004$. This further supports the idea that BL~Lacs
jets have more shocks as higher fractional polarization is expected
for more ordered magnetic fields \citep[e.g.,][]{hughes85}. Similar
results were reported by \cite{aller03} and \cite{lister05}, who
suggested that jets in BL~Lac objects may have more transverse shocks
that increase the polarization, while in quasars the shocks could be
weaker oblique shocks with less polarization. Indeed, \cite{hughes11}
show how many polarization events seen in single-dish radio data can
be modeled with oblique shocks.

If the transverse shocks in BL~Lacs are stronger than the shocks
  in quasars, one would also expect their flux density to be higher. A
  simple test is to compare the flux density of the jet components to
  the flux density of the core. Contrary to the expectation, we find
  that the fraction of the jet to core flux density is higher in
  quasars (mean 0.49) compared to the BL~Lacs (mean 0.20). However, we
  note that the situation is more complex and the simple comparison
  may be affected by differing core properties of quasars and BL~Lacs
  or different Doppler boosting in the core and jet components of
  BL~Lacs and quasars. For example, if the Doppler boosting factors of
  BL~Lacs are smaller than in quasars, as suggested by
  \cite{hovatta09}, the jet to core flux density ratio appears smaller
  than in quasars for same flux densities (assuming continuous
  emission for the core and a moving feature for the jet components). Another possibility is that some difference in the particle acceleration process in BL~Lacs and quasars causes the optically thin spectra to differ.

\subsection{Comparison to $\gamma$-ray data}\label{sect:gamma}
Out of the 190 sources we have studied, 119 are associated with a $\gamma$-ray source in the {\it Fermi} Gamma-ray Space Telescope 1FGL or 2FGL catalogs \citep{abdo10d,nolan12}.
\cite{pushkarev12b} found a statistically significant difference between the jet spectral index in the $\gamma$-ray associated and unassociated sources with the unassociated sources having steeper spectra. We do not find significant differences between the $\gamma$-ray associated and unassociated objects in our sample (K-S test $p=0.07$ for core and $p=0.08$ for jet components). 

The sample of \cite{pushkarev12b} includes 135 $\gamma$-ray associated and 184 unassociated objects, making the fraction of unassociated sources much larger than in our sample. If we compare only those sources that are common in both samples (66 $\gamma$-ray associated and 34 unassociated), we find that the difference between the spectra of the jet components in the associated and unassociated objects is not statistically significant in our data (K-S test $p=0.61$) or in the data of \cite{pushkarev12b} (K-S test $p=0.17$). This indicates that the difference in our results compared to \cite{pushkarev12b} is due to the larger fraction of unassociated sources in their sample.

\subsection{Spectra of individual components}\label{sect:spectra}
With the four separate frequency bands, we can fit for the shape of the spectrum. In order to account for the component size, we calculate an average 
over component size in the image plane. We use the fitted size at 15\,GHz for all frequency bands. As discussed in Sect.~\ref{sect:maps} this could also be done in the 
visibility plane by fitting and cross-identifying the components at the different bands. However, for the reasons mentioned in the same section, 
this is not practical for our large sample. Therefore we have examined a few individual cases in detail to verify that the overall correspondence between the 
spectral shape when using flux densities from cross-identified model components and when using average values extracted in the image plane is very 
good. We only use components that are more than a beam size away from the core component to avoid the contribution from the beam-convolved core.
In some cases, bright nearby jet components can also contribute to the flux density derived from the images. 
Therefore, as in \cite{hovatta12}, we have 
defined isolated jet components to be ones not affected by other nearby bright components. For each component we calculated 
the combined contribution of all the other jet components in the map at the component's peak intensity position. 
If this sum was less than 30\% of the component's total intensity, we considered the component to be isolated. We study the spectral shape in these isolated jet components only. We also exclude components that are less than 10\,mJy in total intensity at any band because these are usually large, diffuse components that can be difficult to cross-identify between the frequency bands. 

The uncertainty in total intensity includes thermal noise (defined in Eq.~\ref{eq:err}), uncertainty due to 
absolute flux calibration (estimated to be 5\% of the total intensity at 15.4, 8.1 and 8.4\,GHz, and 7.5\% at 12.1\,GHz) and an uncertainty due to image alignment, all added in quadrature. The image alignment uncertainty is estimated by shifting the centroid of the component by 1 pixel (0.1~mas) in each direction (near the average image alignment shift in our sample) and calculating the spectral index in each case. The alignment uncertainty is then defined as the standard deviation of these values (see also the discussion in \citet{fromm13} on how to estimate the alignment uncertainty through Monte-Carlo simulations). 

With only four frequency bands we are limited in the fitting process due to the small number of 
data points. Because most of the jet components have an optically thin spectral index between 8.1 and 15.4\,GHz, we start by fitting 
a simple power-law spectrum to the total intensity at each frequency band. We calculate the $\chi^2$ of the fit to determine if this simple 
fit can adequately explain the spectrum. We choose a limit of $\chi^2 < 5.99$, which corresponds to a 95\% confidence limit when there are 
two degrees of freedom. Additionally, we require the fitted spectral index to be less than $0$ to exclude spectra that are inverted. 
The vast majority of the isolated components in our sample (116 out of 125) can be fit 
with a simple power-law. 

\subsubsection{Components with non-power-law spectra}
\begin{figure}[ht!]
%\epsscale{.80}
\begin{center}
\includegraphics[scale=0.35]{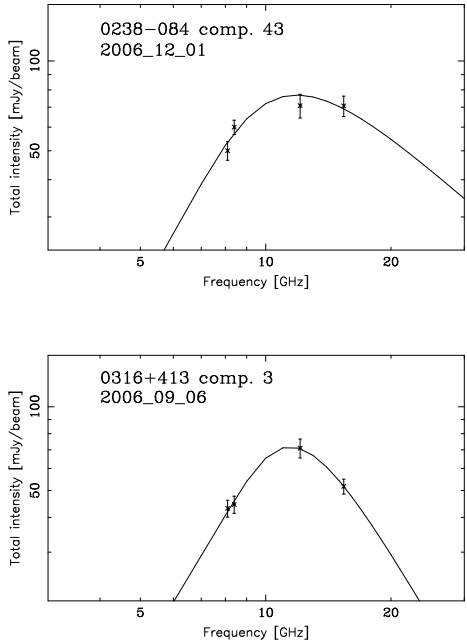}
\caption{Spectra of isolated jet components in the radio galaxies 0238$-$084 (top) and 0316+413 (bottom) that are well represented by a synchrotron spectrum shown as solid line.\label{fig:spectra}}
\end{center}
\end{figure}

In the 9 cases where the criteria for a simple power-law spectrum are not met, we fit a spectrum of a homogeneous synchrotron source of the following form \citep[e.g.,][]{pacholczyk70}
\begin{equation}\label{eq:synch}
I(\nu) = I_m\left(\frac{\nu}{\nu_m}\right)^{5/2}\frac{1-\exp(-\tau_m(\nu/\nu_m)^{\alpha-5/2})}{1-\exp(-\tau_m)},
\end{equation}
where $\nu_m$ is the turnover frequency, $I_m$ is the maximum total intensity reached at the turnover,  $\alpha$ is the optically thin spectral index, 
and $\tau_m$ is the optical depth at the turnover, which can be approximated by
\begin{equation}
\tau_m = \frac{3}{2}\left(\sqrt{1-\frac{8\alpha}{15/2}}-1\right)
\end{equation}
\citep{turler99}. In the above equations we also assume the optically thick spectral index to be 
5/2 because we do not have enough frequency bands to constrain the fit. By visually inspecting the spectra, we exclude seven cases where the 
synchrotron fit is clearly not good. The spectra of the two remaining components, in the radio galaxies 0238$-$084 (NGC~1052) and 0316+413 (3C~84), are shown in 
Fig.~\ref{fig:spectra}. 

These two sources have been studied extensively with a wider frequency coverage by \citet{vermeulen03} and \citet{kadler04} (0238$-$084) and by \cite{vermeulen94} and \cite{walker00} (0316+413). \cite{vermeulen03} studied the pc-scale jet of 0238$-$084 at seven frequency bands between 1.4 and 43\,GHz and found the spectra to be consistent with free-free absorption together with synchrotron self-absorption. Our component 43 is in the western jet of 0238$-$084, which \cite{kadler04}, using data between 5 and 43\,GHz, found to be more consistent with free-free absorption. Similarly, our component 3 of 0316+413 is in the northern jet of the source, where free-free absorption is suggested as the cause for the turnover \citep{vermeulen94,walker00}. Based on the low brightness temperature of the northern jet, \cite{vermeulen94} suggested that the turnover is likely to be free-free absorption. This was confirmed by \cite{walker00}, who used data between 2.3 and 43\,GHz and found the spectral index below the turnover to be 
+4, much steeper than the expected value of $+2.5$ for synchrotron self-absorption. 

The brightness temperature of the jet components can be calculated using Eq.~\ref{eq:Tb}.
To determine the flux density and size of the component at the turnover frequency, 
we have taken the nearest observed frequency band to the turnover frequency and taken the fitted Gaussian component size and flux density at that 
frequency as an estimate of the turnover values. This way we determine the brightness temperature of component 3 in 0316+413 to be $\sim 6\times10^8$\,K. 
This is similar to the estimate given in \cite{vermeulen94}. They conclude that for the turnover to be caused by synchrotron self-absorption, the magnetic field energy density would have to dominate the particle energy density by more than a factor of $10^{14}$, which is unlikely considering that jets are often observed near equipartition \citep{readhead94}. 
For 0238$-$084 component 43 we obtain a brightness temperature of $\sim 2\times10^9$\,K, still low for synchrotron self-absorption.
Even though our limited frequency range and small number of independent frequencies does not allow more detailed modeling of the spectra in the sources in our sample, our results for these two sources are consistent with the findings of the more detailed studies.

\section{Spectral index steepening along the jets}\label{sect:aging}
It is obvious from the difference in the core and jet spectral indices that some steepening in the index is occurring along the jet. 
One needs to be careful in distinguishing instrumental effects, e.g., poorer ({\it u,v}) plane coverage at short baselines at the 15\,GHz band, 
from intrinsic effects, i.e., steepening due to radiation losses (spectral aging). The lack of short baselines at 15\,GHz results in loss of 
sensitivity in the more extended emission in the jet, artificially steepening the spectral index. The effect can be diminished by 
clipping the ({\it u,v})-coverage to be the same in each band as we have done. Furthermore, the frequency difference between our bands 
is only a factor of two, in which case we do not expect significant steepening due to instrumental effects. We have verified this assumption by detailed simulations presented in Appendix~\ref{app:steep}.

\subsection{Spectral steepening observations}
\begin{figure}[ht!]
%\epsscale{.80}
\begin{center}
\includegraphics[scale=0.4]{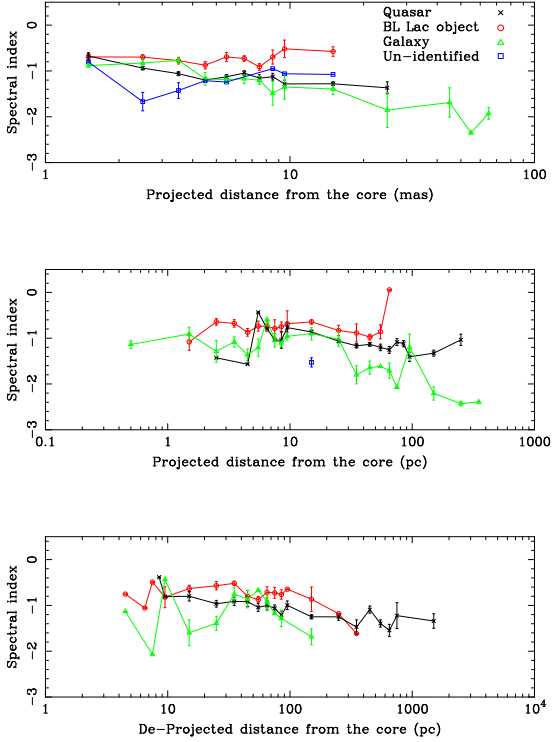}
\caption{Spectral index of all sources along the jet ridge line binned by distance. Top panel shows the projected distance in milliarcseconds, middle panel shows the projected distance in parsecs and the bottom panel the de-projected distance in parsecs. The projected distance is available for sources with a known redshift and the de-projeced distance for sources with an estimate for the viewing angle (see text for details). The high spectral-index point for BL~Lacs in the middle panel is due to a single source, 1803+784, and a noisy patch at the end of its jet. Quasars are shown with crosses, BL~Lacs with circles (red in the online journal), galaxies with triangles (green in the online journal) and optically un-identified objects with squares (blue in the online journal). 
(A color version of this figure is available in the online journal.)\label{fig:alpha_dist}}
\end{center}
\end{figure}

To study the steepening along the jets, 
in Fig.~\ref{fig:alpha_dist} we show the spectral index as a function of distance along the jet ridge line. 
Following \cite{laing13}, we first bin the ridge lines of each individual source in bins the size of the beam along the ridge line. We then bin these in projected distance to clarify the plot but the statistical tests are done using the data from individual sources and all beam-size binned values. When we look at the projected distance from the core in milliarcseconds (top) or parsecs (middle), the quasars show a significant negative correlation (Kendall's $\tau_K = -0.22$, $p=1\times10^{-14}$) while in BL~Lacs the correlation is not significant ($\tau_K = 0.01$, $p=0.81$). However, when we take the viewing angle of the source into account 
and calculate the de-projected distance in parsecs\footnote{The viewing angles $\theta = \tan^{-1}[(2\beta_\mathrm{app})/(\beta_\mathrm{app}^2+\mathrm{D_{var}}^2-1)]$ 
are determined with Doppler factors D$_\mathrm{var}$ from \cite{hovatta09} and maximum apparent speeds $\beta_\mathrm{app}$ 
from Paper X.}, both quasars and BL~Lacs show a significant (quasar $p=3.1\times10^{-10}$, BL~Lac $p=0.010$) 
negative correlation ($\tau_\mathrm{quasar} = -0.26$, $\tau_\mathrm{BL Lac} = -0.20$). This indicates that the intrinsic steepening of the 
spectral indices is similar in the two classes of object, despite the flatter overall spectra of the BL~Lacs.

\begin{figure*}[ht!]
%\epsscale{.80}
\begin{center}
\includegraphics[angle=-90,scale=0.6]{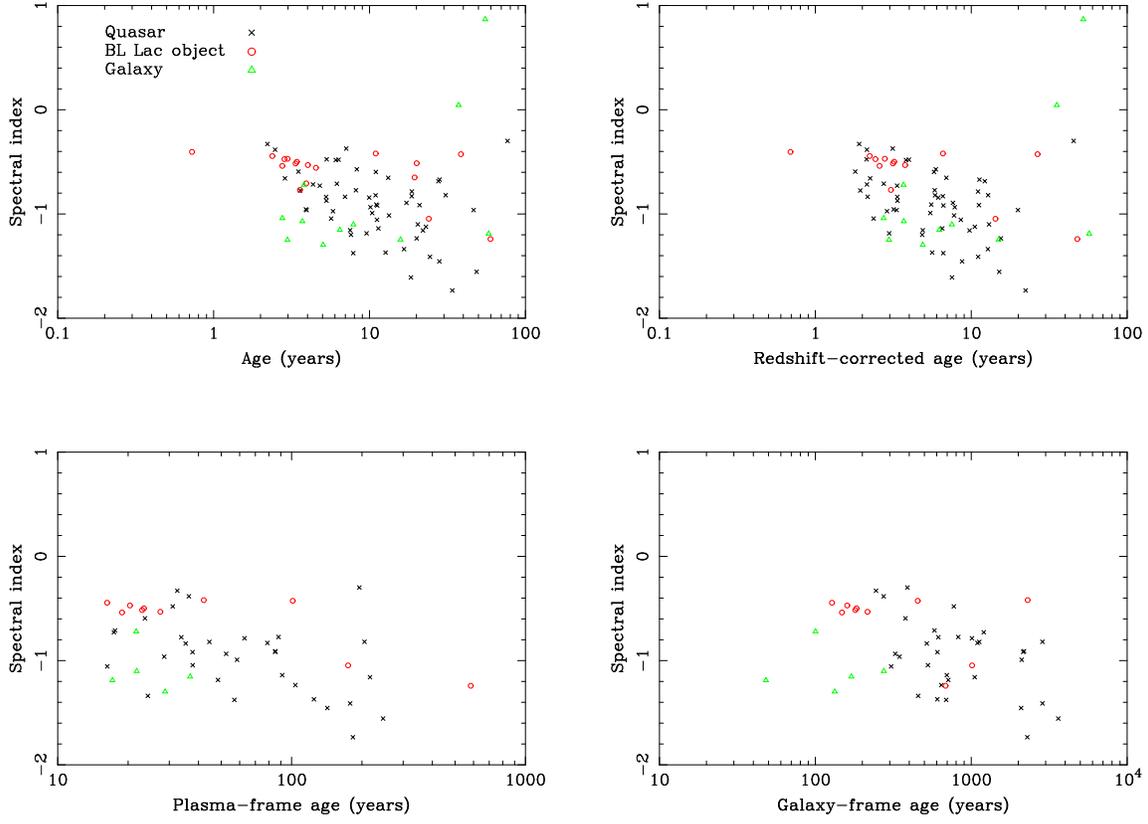}
\caption{Spectral index in the jet components against the age of the component. Top left panel shows the age in observer's frame, top right is redshift-corrected, bottom left shows the age in the plasma frame and bottom right in the host galaxy frame. Quasars are shown with crosses, BL~Lac objects with circles (red in the online journal) and galaxies with triangles (green in the online journal). Note the different scaling of the x-axis in each panel.
(A color version of this figure is available in the online journal.)\label{fig:age}}
\end{center}
\end{figure*}

In the top left panel of Fig.~\ref{fig:age}, 
we compare the age of the jet components to the spectral index. The ages are calculated using the ejection epoch determined over 
several years of observations (Paper X) and are listed in Table~\ref{comptable}. These are available for 89 jet components 
in our sample that are more than a beam size away from the core. As expected for cooling, older components have steeper spectra ($\tau_K=-0.26$, $p=0.0004$). This trend seems to be 
more significant in quasars ($\tau_K=-0.31$, $p=0.0007$) than in BL~Lacs ($\tau_K=-0.35$, p$=0.048$).

In order to properly account for cosmological and relativistic effects, in the top right panel we show the age corrected for redshift by 
$t_z = t_\mathrm{obs}/(1+z)$. Due to the fairly narrow range of redshifts in our sample, this has only a minor effect on the correlations. 
The bottom panels of Fig.~\ref{fig:age}  show the correlation when relativistic effects are accounted for. To translate the time into the 
jet plasma frame we have multiplied the redshift-corrected time by the Doppler factor for each source, taken from \cite{hovatta09}, i.e., 
$t_p = D_\mathrm{var}t_z$. Doppler factors were available for 69 jet components in our sample. The correlation coefficients are similar but the significances are reduced, most likely due to the smaller number of components. The bottom right 
panel shows the age of the component in the host galaxy frame which can be obtained by multiplying the age in the plasma frame by the 
Lorentz factor of the jet, $t_h = \Gamma t_p$. The Lorentz factors are calculated using the Doppler factors and apparent speeds of the fastest 
component in the source (Paper X). The number of components is the same as in the plasma frame case but the 
correlation is less significant in quasars while in BL~Lacs it is no longer significant ($\tau_\mathrm{quasar} = -0.25$, $p=0.044$, $\tau_\mathrm{BL Lac} = -0.07$, $p=0.8$). 
There are only 10 jet components available for the BL~Lacs, and the lack of correlation could be simply due to small number statistics. However, it is clear that in at least quasars we see spectral index steepening that is related to the age of the component.

The ages of the components can also be estimated from their
  apparent speed, as was done in Paper X, where we studied the
  correlation between the apparent speed of the components as a
  function of the projected distance along the jet (see Fig. 12 in
  Paper X). In Paper X, we argued that the bottom right corner of the
  figure is undersampled because the components have faded below the
  detection threshold of MOJAVE. This is supported by our spectral
  index analysis where we find that all components older than 250
  years, as calculated from the speeds, have spectra steeper than $-0.5$.

\begin{figure}[ht!]
%\epsscale{.80}
\begin{center}
\includegraphics[angle=-90,scale=0.3]{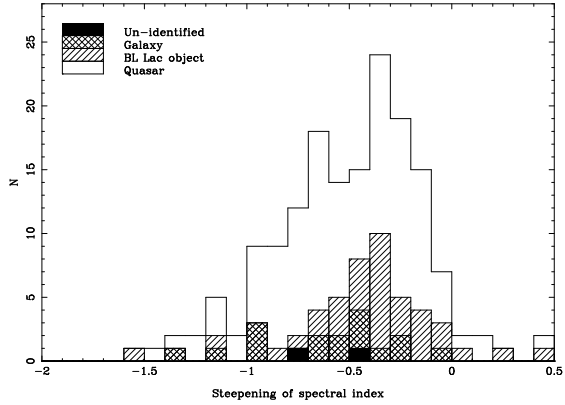}
\caption{Distribution of spectral index steepening $\Delta\alpha$ in the jets.\label{fig:steepening}}
\end{center}
\end{figure}

We can also look at the spectral steepening in the jets by comparing the spectral index at the edge of the convolved core with the median jet ridge line spectral index. We do the comparison this way because due to the flattening of the spectra at the component locations, the steepening is not well-described by a linear power-law fit. The edge of the convolved core (shown as dashed vertical line in Fig.~\ref{spmap}) is defined as the full-width half maximum of the Gaussian resulting from the convolution of the Gaussian core component with the Gaussian beam. The spectral index at the edge of the convolved core is estimated by linearly interpolating between the spectral index at the last ridge line point within the convolved core and the first ridge line point beyond the convolved core. In this manner we can estimate the steepening, defined as $\Delta\alpha = \alpha_\mathrm{jet-median} - \alpha_\mathrm{core-edge}$ in 162 cases. The distribution divided into optical sub-classes is shown in Fig.~\ref{fig:steepening}. 
The spectral index values in quasars steepen on average by $\Delta\alpha = -0.52\pm0.03$ and in BL~Lacs by $\Delta\alpha = -0.39\pm0.06$. According to the K-S test the probability for them to come from the same population is $p=0.029$.

\begin{figure}[ht!]
%\epsscale{.80}
\begin{center}
\includegraphics[scale=0.45]{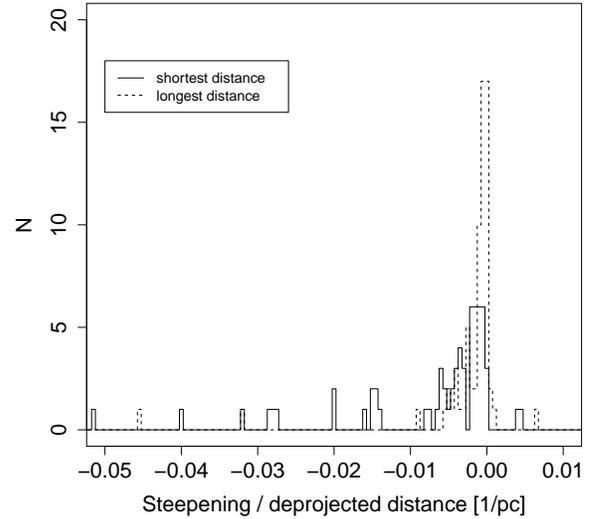}
\caption{Distribution of the spectral index steepening per deprojected parsec in the two extreme cases where the steepening is estimated to occur at the first point where the ridge line spectral index crosses the median spectral index value (shortest distance, solid line) and where the steepening occurs over the entire jet length (longest distance, dashed line). The plot contains 90\% of the components with some extreme outliers excluded for clarity.\label{fig:steep_parsec}}
\end{center}
\end{figure}

We use the median value of the spectral index as the final spectral index value because it is less affected by noise and ({\it u,v})-coverage effects. Therefore it is not straightforward to estimate the distance over which the steepening occurs. Instead, we define a range of distances to estimate the typical steepening per parsec values.
First, we calculate the distance between the edge of the convolved core and the first point where the jet spectral index crosses the median value. This gives us a lower limit to the distance. As the other extreme, we use the length of the jet in the ridge line analysis. For the sources with known redshift, Doppler factor and viewing angle, we can estimate how much the spectral index steepens per deprojected parsec. Figure~\ref{fig:steep_parsec} shows the distributions for the two extreme cases. We find that the median steepening is $\sim -1\times10^{-3}$~pc$^{-1}$ if the steepening occurs over the entire jet length and $\sim -4\times10^{-3}$~pc$^{-1}$ if the steepening occurs at the lower limit distance. 

Another interesting question is whether stationary components in
  the jets show flatter spectra and affect the steepening we
  observe. There are 40 components identified as stationary in our
  data set but only five of those are more than a beam width away from
  the core. Therefore higher resolution observations are required to
  study this further.

\subsection{The physical cause of spectral steepening}
\subsubsection{Radiative losses}
Here we discuss the steepening in the synchrotron spectrum that can be caused by synchrotron losses.  If nonthermal particles with an energy spectrum $Q_{\rm inj}=\gamma^{-p}$ for $\gamma_{\rm min} < \gamma <\gamma_{\rm max}$ and $Q_{\rm inj}=0$ elsewhere are constantly injected into a synchrotron emitting region, then a synchrotron break in the electron spectrum can occur such that $N\propto \gamma^{-p}$ for $\gamma_{\rm min} <\gamma<\gamma_{\rm br}$ and $N\propto \gamma^{-(p+1)}$ for $\gamma_{\rm br}<\gamma<\gamma_{\rm max}$, which produces a corresponding break in the synchrotron spectrum of $\Delta \alpha = -0.5$ \citep{kardashev62}.  For self-similarly expanding emission regions such that the transverse radius of the jet R depends on the distance from the jet vertex $z$ as $R\propto z^a$, the magnetic field will decay as a power law: $B'(t')=B_0'(t'/t_0')^{-a}$, where $a$ depends on the magnetic field geometry and primed quantities refer to the comoving frame.  The break energy $\gamma_{\rm br}$ can be estimated by solving $t'=\gamma_{\rm br}(d\gamma/dt'(\gamma_{\rm br}))^{-1}_{\rm syn}$ for $\gamma_{\rm br}$, where $(d\gamma/dt')_{\rm syn}=-b\gamma^2(t'/t_0')^{-2a}$ is the synchrotron loss equation.  This yields
\begin{align}
\gamma_{\rm br}\approx b^{-1}t'^{2a-1}t_0'^{-2a}, \quad b=\frac{\sigma_TB_0'^2}{6\pi m_e c},\label{eq:loss}
\end{align}
which we spot checked numerically by solving the continuity equation \citep{ginzburg64} that governs the time evolution of $N(\gamma)$ in an expanding (conical) jet with constant injection, adiabatic losses, and synchrotron losses \citep[e.g.,][]{gupta06,fromm13}.  Note that for $a=2$, \citet{kardashev62} finds that $\gamma_{\rm br}\propto t^{3}$ for the same scenario we consider, in agreement with our above scaling.  The synchrotron break frequency can now be expressed as $\nu_{\rm br}\approx \gamma_{\rm br}^2 eB/(m_ec)$, implying that $\nu_{\rm br}\propto t'^{3a-2}$ \citep[eqn.~A3,][]{marscher85}.  Thus, for components' spectra to steepen as they move down the jet as required by our data, then $a$ must be $<2/3$. Note that for a toroidally dominated magnetic field where $B' \propto R_{\perp}^{-1}$, $a<1$ refers to a collimating jet, $a=1$ to a conical jet, and $a>1$ to a decollimating jet. 

All of the above analysis applies for cooling due to inverse Compton (IC) emission as well, since its cooling function also depends on $E^2$. The principal difference is that the seed photon energy density in IC emission plays the role of $B^2$ in synchrotron emission.  However, except for the cosmic microwave background (CMB), seed photons from the broad line region, black hole accretion disk, or the emitting region itself (as in synchrotron self-Compton emission), the seed energy density will decay in a manner equivalent to $a=1$ or higher.  The CMB's energy density is low enough that it will not produce breaks in the GHz-range spectra.  

While there is some evidence that the jets on parsecs scales appear conical with intrinsic opening angles that scale inversely with the Lorentz factor \citep{jorstad05, pushkarev09, clausen-brown13}, it is possible that the derived opening angle values are affected by the components at any given epoch only illuminating a small fraction of the jet (Paper X).
Thus, we cannot exclude the possibility of a collimating jet (i.e.,
$a<1$) that would potentially allow for radiative losses alone to explain the spectral steepening we observe.

\subsubsection{The time evolution of $\gamma_{\rm max}$}
Another possibility is that we are not seeing a synchrotron cooling break, but instead we are observing the time evolution of $\gamma_{\rm max}$, the high energy cutoff of the electron energy spectrum, which produces an exponential cutoff in the synchrotron spectrum.  If no injection occurs, then $\gamma_{\rm max}$ decreases with time through adiabatic and radiative losses, producing an cutoff in the synchrotron spectrum with arbitrarily high $|\Delta \alpha|$.  This type of scenario could occur if the conditions for particle acceleration in jets are only intermittently realized, such as in recollimation shocks or instabilities.

The expected spectral break $\Delta\alpha$ is derived in Appendix \ref{app:deriv}, where we find that if the initial observed spectral index closer to the radio core is $\alpha(z_{\rm obs})=\alpha_0$, then 
\begin{equation}
\begin{split}
\Delta \alpha &= \alpha(z_f) - \alpha(z_{\rm obs}) = -\frac{\nu}{\nu_{\rm cut,f}} \\
&\approx -0.53
  \left(\frac{B_0'}{\text{1\,G}}\right)^{3}\left(\frac{z_{\rm f,pc}}{300}\right)^{-1}\left(\frac{\Gamma}{10}\right)^{-2}\left(\frac{\delta}{10}\right)^{-1}\\
&\times\left(\frac{R_{\rm exp}A}{4\times7}\right)^{10/3}\left(\frac{\nu_{obs}}{10\text{\,GHz}}\right),
\end{split}
\end{equation}
where $\alpha(z_f)$ is the spectral index value we can measure at the end of the jet, $B_0'$ is the comoving magnetic field strength (typically $B_0' \approx 1$\,G at 1\,pc \citealt{pushkarev12}), $\Gamma$ is the jet Lorentz factor, $\delta$ the jet Doppler factor, $R_{\rm exp}$ is the expansion factor of the components, and $A=z_{\rm obs}/z_i$ is the ratio of the location in the jet where injection stops to where the component is first observed.
For typical jet parameter values, this would then produce an average gradient in spectral index of $\Delta \alpha \approx 0.5/\text{length [pc]} = ((1-1/R_{\rm exp}) 300\text{\, pc})^{-1}\sim 2\times 10^{-3}$\,pc$^{-1}$.  Note that the location where particle injection stops in our parametrization here is $z_i=z_{\rm f,pc}/(R_{\rm exp}A)$, and for the above fiducial values of $A$ and $R_{\rm exp}$ we find that $z_i \sim 10$\,pc.

\begin{figure}[ht!]
%\epsscale{.80}
\begin{center}
\includegraphics[scale=0.45]{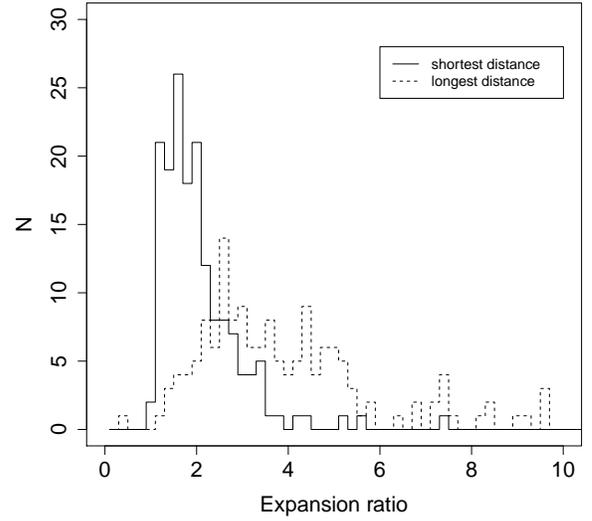}
\caption{Distribution of the component expansion ratio between the size at the edge of the convolved core and the two extreme cases where the sizes are estimated at the first point where the ridge line spectral index crosses the median spectral index value (shortest distance, solid line) and at the jet length (longest distance, dashed line).\label{fig:expansion}}
\end{center}
\end{figure}

Hence our proposed mechanism can naturally explain how components near the core with optically thin spectral indices can steepen by $\Delta \alpha = -0.5$, as long as they expand by a moderate factor.  We can study the typical amount of expansion in the jets by estimating the sizes of the components at various locations in the jet. To do this, we use all the components for each source in all the epochs in Paper X. We then fit a power law for the relation between the component distance from the core and the size of the component to get an estimate of the typical size at a given distance. Fig.~\ref{fig:expansion} show distributions for the expansion ratio between the size at the edge of the convolved core and the two extreme distances at the point where the ridge line spectral index first crosses the median spectral index value and at the total jet length. In most of the sources the components expand by a factor of 1.5 to 4, depending on the distance estimate, with a median value of 2.4 for the combined distributions, in accordance with the above calculations.

Of course, this explanation rests on the critical assumption that
nonthermal particle acceleration (i.e., particle injection) is no
longer occurring in these components.  We also note that we do not
know the functional form of the electron spectrum's high energy
cutoff; if it terminates more gradually near $\gamma_{\rm max}$ rather
than the sudden step function cutoff we assume, then this would reduce
our estimate of $\Delta \alpha$ and possibly decrease the sensitive
dependence $\Delta \alpha$ has on model parameters such as $B'_0$ and
$R_{\rm exp}$.  If this steepening mechanism is true, this suggests
that particle acceleration is not a continuous process occurring
throughout jets, but rather an intermittent one that occurs close to
the SMBH and at other locations in the jet as a whole, such as in
X-ray emitting knots on the arcsecond scales where the synchrotron lifetime is short.

Another cause for spectral breaks of $\Delta \alpha = -0.5$ can be derived if the sink term $N/t_{\rm esc}$ is added to the continuity equation, in which case a spectral break can form due to synchrotron cooling and particle escape.  As a component's age $t'$ approaches infinity, an electron spectral break exists and scales as $\gamma_{\rm br}\propto t_{esc}^{-1}$ \citep{dermer09}.  This translates into an observed spectral break frequency for an expanding component that scales as $\nu_{\rm br}\propto t'$ for a conical jet where $a=1$ and $t_{\rm esc}\propto R_{\perp}$, and thus cannot be the cause of the spectral steepening that we observe.

Furthermore the spectral steepening could be caused by the evolution of the electron power-law index $p$, which could be due to the time evolution of the distribution of the pre-accelerated particles, of the acceleration process itself, or of some other acceleration related process.  However, the cause is more likely synchrotron cooling or the time-evolution of $\gamma_{\rm max}$ since these processes are more clearly related to component aging and expansion.

\section{Conclusions}\label{sect:conc}
We have studied the spectral distributions of 190 parsec-scale AGN jets using VLBA data at 8.1, 8.4, 12.1 and 15.4\,GHz. We provide spectral index maps between 8.1 and 15.4\,GHz for all the sources and study the spectral index evolution along the jets. Our findings can be summarized as follows:
\begin{enumerate}
\item The spectral index of the unresolved jet core is similar in all optical classes, and the mean value for our sample is $\alpha_{core}=0.22\pm0.03$.
\item The jet spectral index of quasars (mean
  $\alpha_{jet,quasar}=-1.1\pm0.04$) is significantly steeper than in
  the BL~Lac objects (mean $\alpha_{jet,BLLacs}=-0.80\pm0.06$). This can be explained if BL~Lac jets have more active sites of particle acceleration, perhaps due to differences in the number or types of shocks compared to quasar jets.
\item We find the spectra to flatten significantly at component locations on average by $\sim0.2$, indicative of particle acceleration or density enhancements in the jet. 
\item There is a significant negative correlation between the component spectral index and the direction of the polarization angle in the jet. Components with polarization parallel to the jet (magnetic field perpendicular to the jet) have flatter spectra, as might be expected for transverse shocks which order the magnetic field and re-accelerate the particles.
\item The jet spectra steepen significantly, on average by $\sim
  -0.45$, over the jet length. This can be due to radiative losses if
  the jets are collimating or due to the evolution of the high-energy
  cutoff from both adiabatic and radiative losses in the electron distribution if the jets are conical. Spectral evolution in some components is confirmed by a significant correlation between the age of the component and the spectral index.
\end{enumerate}

\acknowledgments
We thank the referee for valuable suggestions that
significantly improved the paper. 
The authors wish to thank Philip Hughes, Ken Kellermann, Dave Meier, Eduardo Ros, and the rest of the MOJAVE team for useful discussions. T. Hovatta thanks Joni Tammi for helpful discussions. Large part of this work was done when T. Hovatta was working at Purdue University.
The MOJAVE project is supported under National Science Foundation grant AST-
0807860 and NASA Fermi grants NNX08AV67G and 11-Fermi11-0019. Work at UMRAO has been supported by
NASA Fermi GI grants NNX09AU16G, NNX10AP16G, NNX11AO13G,
and NNX13AP18G, National Science Foundation grant AST-0607523, and by funds
for operation from the University of Michigan. T. Hovatta was supported in part by the
Jenny and Antti Wihuri foundation and by the Academy of Finland project number 267324. D. Homan was funded by National Science Foundation grant AST-0707693. 
Part of this work was done when T. Savolainen and Y. Y. Kovalev were research 
fellows of the Alexander von Humboldt Foundation. Y. Y. Kovalev was supported in part by the Russian Foundation 
for Basic Research (grant 13-02-12103.) and the Dynasty Foundation. A.~B.~Pushkarev was supported by the ``Non-stationary processes
in the Universe'' Program of the Presidium of the Russian Academy
of Sciences. This paper is based on VLBA data, project codes BL137A-L.
The National Radio Astronomy Observatory is a facility
of the National Science Foundation operated by under cooperative
agreement by Associated Universities, Inc. This research has made use of NASA's Astrophysics Data System, and the NASA/IPAC Extragalactic Database (NED). 
The latter is operated by the Jet Propulsion Laboratory, California Institute of Technology, under contract with the National 
Aeronautics and Space Administration.

{\it Facilities:} \facility{VLBA} \facility{UMRAO}

\appendix
\section{Effect of ({\it u,v})-coverage on spectral index steepening along the jet}\label{app:steep}
\begin{figure}[!ht]
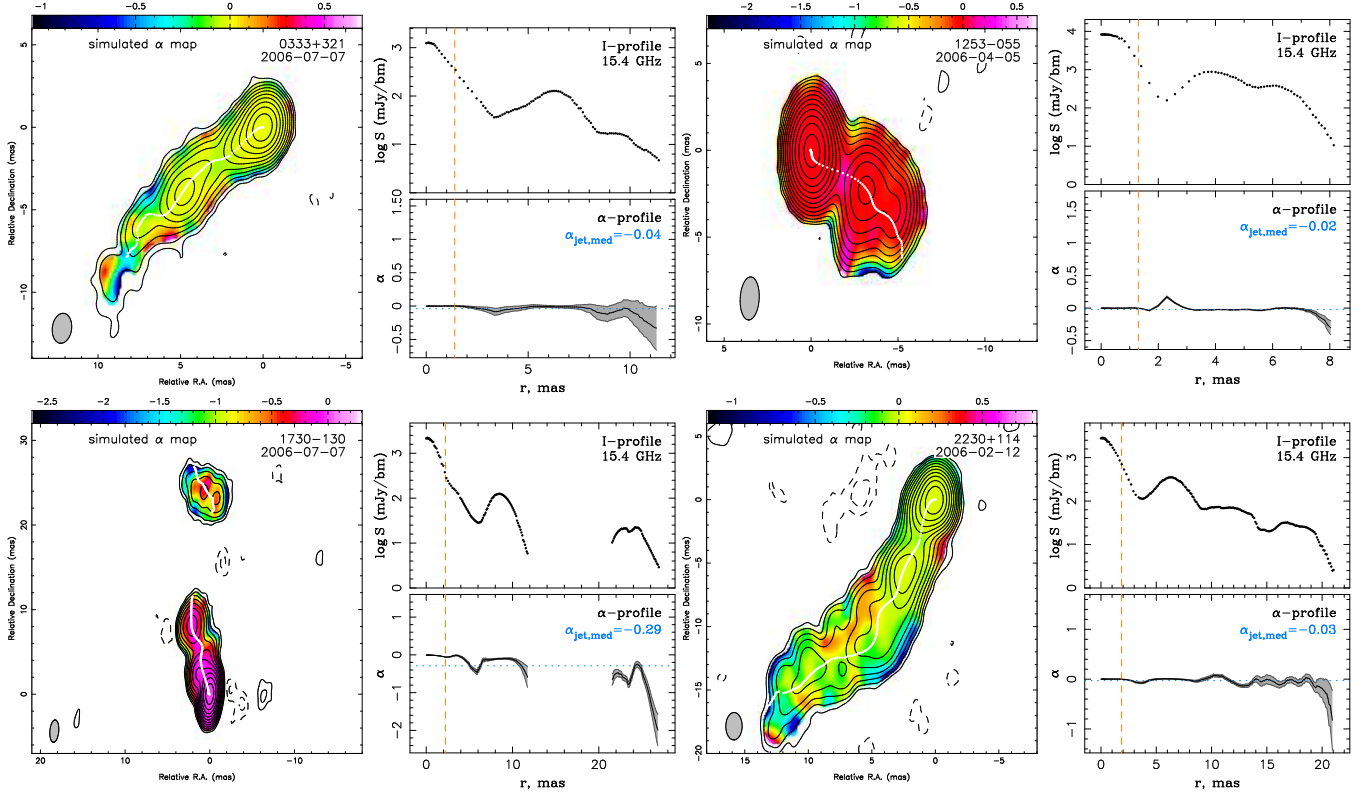

\begin{center}
\includegraphics[scale=0.35,angle=-90]{f13a.ps}
\includegraphics[scale=0.35,angle=-90]{f13b.ps}
\includegraphics[scale=0.35,angle=-90]{f13c.ps}
\includegraphics[scale=0.35,angle=-90]{f13d.ps}
\caption{Simulated spectral index maps of four sources. The dashed vertical line (orange in the online journal) indicates the edge of the convolved core. The dotted horizontal line (blue in the online journal) shows the median jet spectral index along the ridge line.
(A color version of this figure is available in the online journal.)\label{steepsim}}
\end{center}
\end{figure}

Most of the jets in our sample show steepening of the spectral index along the jet. In order to verify that the steepening is intrinsic to the source we used simulations to examine the effect of non-identical ({\it u,v}) - coverage between the 8.1 and 15.4\,GHz maps. The simulations were carried out in several steps:
\begin{enumerate}
\item A Stokes I model of a real source was created from calibrated ({\it u,v}) data at 8.1\,GHz using Difmap. 
\item The original ({\it u,v}) data at 8.1 and 15.4\,GHz were loaded into AIPS and the task UVMOD was used to replace the real data with 
the values produced in the previous step. Additionally, random noise of the same order as 
seen in our real data was added. This step produced maps with the same intensity but differing ({\it u,v}) - coverage for the two bands.
\item The simulated ({\it u,v}) data were then imaged in Difmap following the same procedure 
as for the real data.
\item  The rms in each image was obtained by shifting the map by 1 arcsec and calculating the rms using the 'imstat' command in Difmap.
\end{enumerate}
We repeated the simulations for 16 jets with different declinations and jet directions to account for variations in the ({\it u,v}) - coverage. Figure~\ref{steepsim} shows four example jets and their ridge line spectral index values extracted from the simulated maps. The typical steepening is best quantified by looking at the median spectral index value of the simulated ridge lines. The mean value for the 16 simulated jets is $-0.1$, and so it is clear that the steepening due to ({\it u,v}) - coverage is minimal compared to the actual jet spectral index values in our sample. The worst case is the source 1730$-$130, where the median jet spectral index from the simulations is $\alpha_{sim}=-0.29$, with values of $\alpha_{sim}\sim-2$ towards the end of the jet. However, the median spectral index is fairly robust against the steepening and even if we do not account for the spectral index values beyond 20 mas from the core where the simulations start to show a clear effect, the median changes only by $-0.09$ to $\alpha_{sim,med}=-0.20$. Therefore we conclude that our analyses are not greatly affected by the ({\it u,v}) - coverage effects. This is not surprising considering that our frequency bands differ only by a factor of two. Any studies using a wider frequency range should take possible ({\it u,v}) - coverage effects into account. This agrees with the simulations of \citet{fromm13} (see their Appendix B), who show that when the frequency range is less than a factor of four, the ({\it u,v}) - coverage mainly affects the edges of the spectral distribution.

\section{Effect of image alignment on the spectral index maps}\label{app:align}
The 2D cross-correlation algorithm is sensitive to spectral index gradients along the jet, and can result 
in wrong alignment between the images, affecting the final spectral index results.
This effect depends strongly on the strength of the gradient and the appearance of the jet with featureless, straight 
jets being more sensitive. The effect of spectral index gradient on the alignment was studied with detailed simulations 
by \citet{pushkarev12} and we only give a summary here (see also the discussion of the method in \citet{fromm13}). We find that in jets with knotty structure and bends the 
2D cross-correlation method is fairly robust, even if a large spectral index gradient is present (errors less than 10\% of the 
beam size). In featureless and straight jets the performance is much worse, with errors up to 27\% of the beam size.
This can result in significant steepening or flattening of the
spectral index, especially at the base of the jet. In order to test if
this depends on the source type, we used simple measures to estimate
how knotty and straight the BL~Lacs and quasars are. To estimate how
knotty the jet is, we calculated the fraction of isolated jet
components with respect to all components in each jet, and to estimate how straight the jet is, we calculated the maximum position angle separation of the jet components. We do not see significant differences between the BL~Lacs and quasars.

\begin{figure}[!ht]
\begin{center}
\includegraphics[scale=0.5]{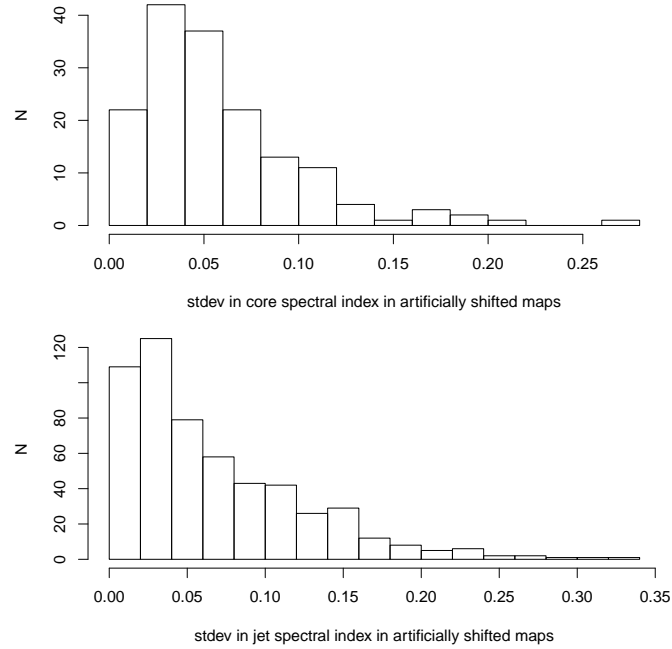}
\caption{Distribution of the standard deviations in spectral index in core and jet components of the artificially shifted maps.}\label{std_dist}
\end{center}
\end{figure}

Another possible caveat is that the 2D cross-correlation method is somewhat subjective, as the user has to decide which alignment to accept. 
This has a large effect, especially in the core region where the spectral index gradient is often large. For example, a shift of the image by just one pixel (0.1~mas) changes the spectral index of a component 1 mas from the core in the source 0648$-$165 by $-1.07$. Because not all the maps shifted by one pixel are considered as acceptable by the user, we cannot simply determine the shifting error by blindly calculating a deviation if we shift the image by one pixel. To reduce individual bias, three people examined the data independently and selected shifts they deemed acceptable. This enabled us to calculate typical expected deviations in the spectral index maps of various sources. In Fig.~\ref{std_dist} we show the standard deviation in the spectral index for core and jet components. The median of both distributions is small, 0.05, indicating that in the majority of cases we do not expect the alignment to cause significant errors in the derived spectral index values. 

\begin{figure}[!ht]
\begin{center}
\plottwo{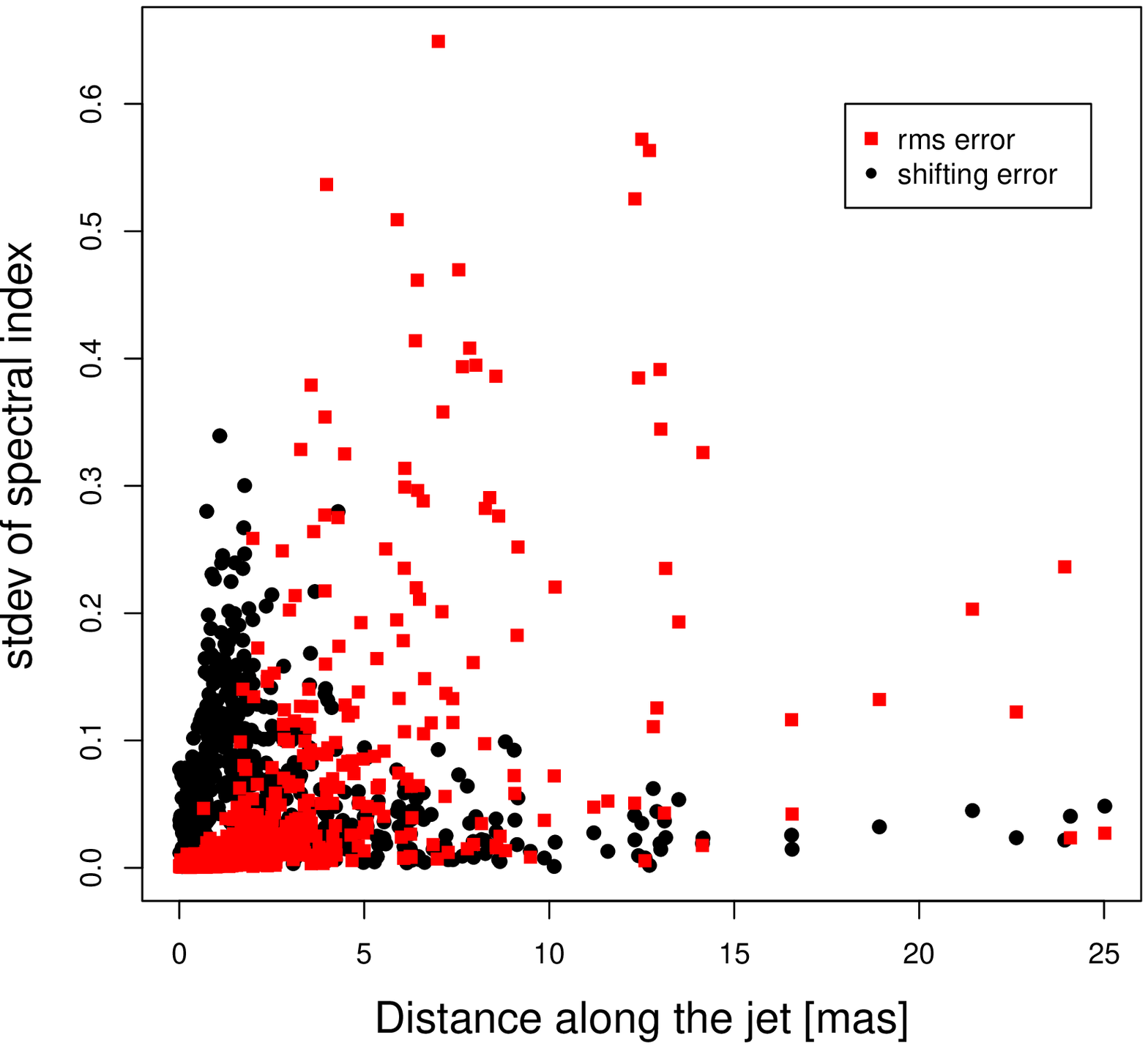}{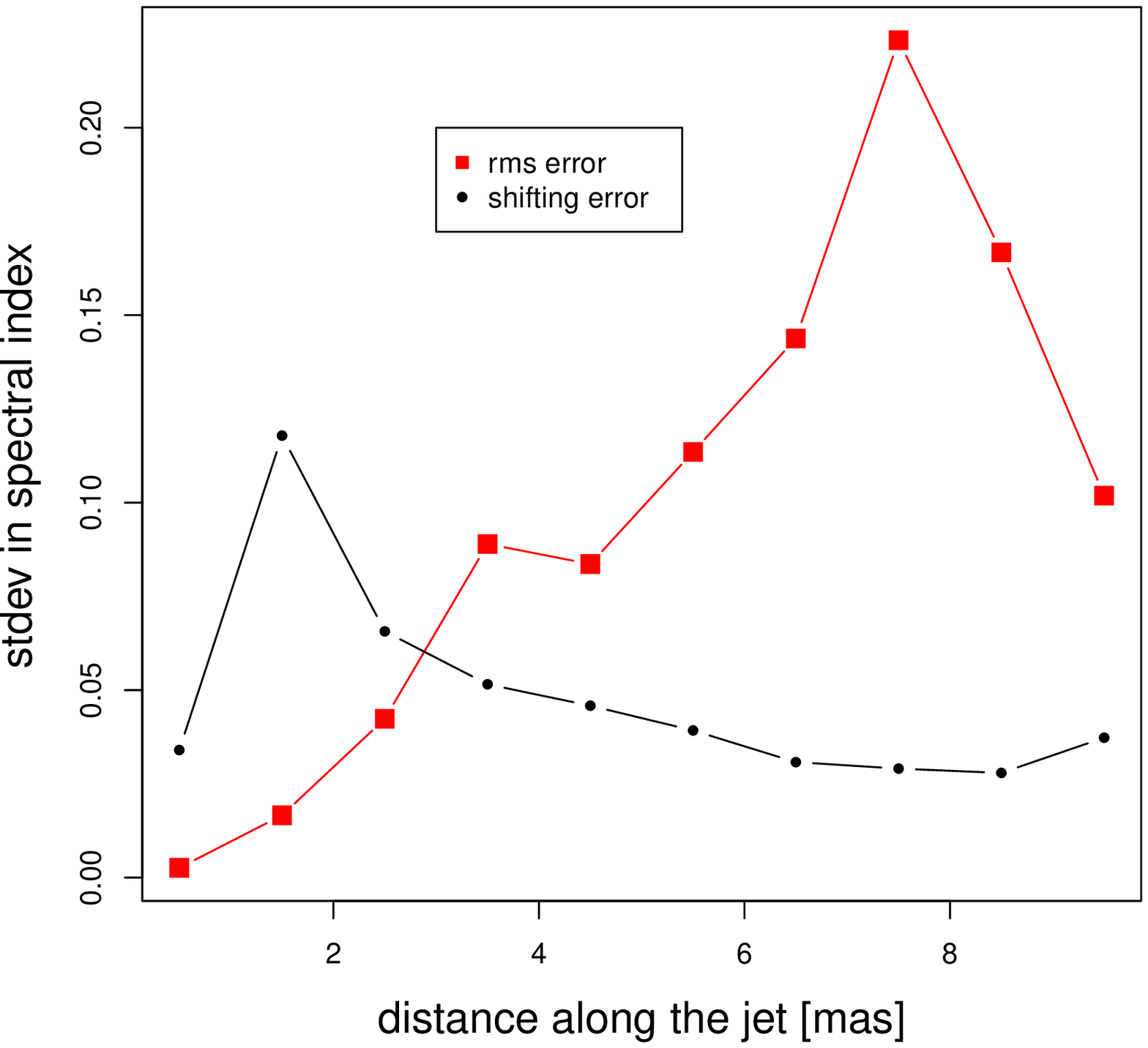}
\caption{Left: Standard deviation for component spectral index from the differently aligned maps as a function of distance from the core, along with the thermal noise (rms error). Right: Same but binned and shown only up to 10~mas from the core.
(A color version of this figure is available in the online journal.)}\label{std_distance}
\end{center}
\end{figure}

The median value is not necessarily representative of the overall jet
because there is a clear dependence between the alignment error and the component distance from the core. We also want to determine at which distance the alignment uncertainty dominates over the thermal noise. Fig.~\ref{std_distance} shows the alignment uncertainty and thermal noise as a function of distance from the core. The binned version indicates that the alignment uncertainty dominates to $\sim$3 mas from the core, after which the alignment uncertainty is near the median value of 0.05. Closer to the core the standard deviation can be much larger. Based on these tests we conclude that the uncertainty induced by the 2D cross-correlation method is small in the majority of the cases but should be investigated in detail when conclusions about core spectral index distribution in individual sources are made.

\section{Effect of beam convolution on the core regions}\label{app:core}
Many of the sources in our sample show the core spectral index as more inverted upstream of the core component and then display a smooth transition to optically thin downstream towards the jet. Some of this effect is due to the beam convolution of the core and jet regions, but the more inverted spectra upstream of the core component can also be due to synchrotron self-absorption. We perform simulations to investigate this effect using a similar approach as in Appendix~\ref{app:steep}. Instead of generating simulated data using CLEAN in difmap, we replace the model in step 1) with Gaussian components by fitting the 15.4\,GHz data in difmap. We then set the model flux densities in the two bands so that we obtain a flat or inverted spectrum in the core and a steep spectrum in the jet. 
\begin{figure}[!ht]
\begin{center}
\includegraphics[scale=0.4,angle=-90]{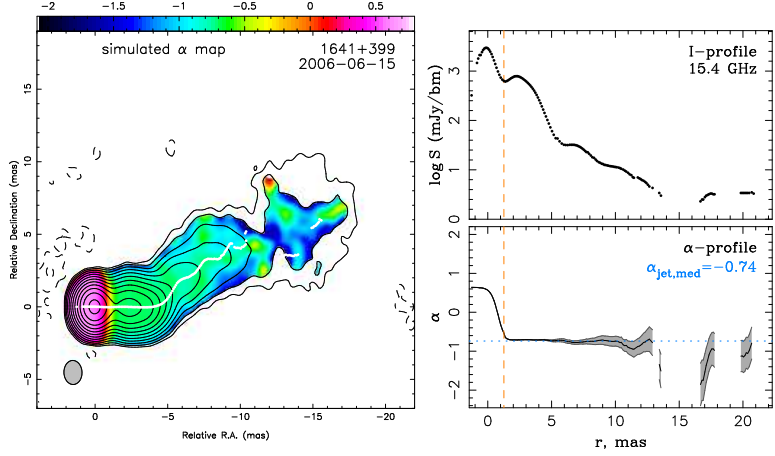}
\caption{Simulated spectral index maps of 1641+399 where the core component spectral index has been set to +0.63 and the jet spectral index to $-0.70$. The dashed vertical line (orange in the online journal) indicates the edge of the convolved core, positioned at 0 distance on the ridge line. The dotted horizontal line (blue in the online journal) shows the median jet spectral index along the ridge line.
(A color version of this figure is available in the online journal.)}\label{coresim}
\end{center}
\end{figure}

In Fig.~\ref{coresim} we show the simulated map and ridge line for the source 1641+399 (3C~345). We set the flux densities of the core components to produce a spectral index of $\alpha_{core}=+0.63$ in the core and $\alpha_{jet}=-0.70$ in the jet components. The orange dashed line in the ridge line plot shows the edge of the convolved core and position 0 in the x-axis the core component position. As can be seen, the spectral index upstream of the core retains the value $+0.63$. Therefore we conclude that the more inverted spectra upstream of the core in our real observations are due to synchrotron self-absorption instead of effects of the beam convolution. Another notable thing is that by the edge of the convolved core, the spectral index has reached the expected jet value, justifying our choice to only use jet components beyond the convolved core region in our analysis of the jet spectral index distributions.

\section{Derivation of $\Delta \alpha$}\label{app:deriv}
The time evolution of $\gamma_{\rm max}$ is governed by the differential equation 
\begin{align}
\dot{\gamma}=-m\gamma/t'-b\gamma^2(t'/t_0')^{-2a},
\end{align}
where $m=2a/3$ and $b$ is defined in Eq.~\ref{eq:loss} \citep[e.g.,][]{gupta06}.  This differential equation describes how $\gamma$ decreases in the comoving frame of the jet with adiabatic (first term) and synchrotron losses (last term) in an expanding jet for which $R_{\perp}\propto z^a$, where $z$ is the distance down the jet. (Since we use the parameter $a$ for both the jet geometry and magnetic field decay, we are implicitly assuming hereafter that $B'\propto R_{\perp}^{-1}$.)  If we assume that particle injection stops occurring in a jet component at $t_i'$, then the $\gamma_{\rm max}$ will initially decrease rapidly due to synchrotron losses, and then more gradually due to adiabatic losses.  If the particle injection is stopped at $t'=t_i'$, such that a particle with random Lorentz factor $\gamma(t'=t_i')=\gamma_i$, and where the magnetic field decays as described above, then
\begin{align}
\gamma(t',\gamma_i,t_i')&=\frac{t'^{-m}}{\gamma_i^{-1}t_i'^{-m}+\frac{bt_0'^{2a}}{1-m-2a}\left(t'^{1-m-2a}-t_i'^{1-m-2a}\right)} \\
&\approx \frac{5}{3bt_i'}\left(\frac{t_i'}{t_0'}\right)^{2}\left(\frac{t'}{t_i'}\right)^{-2/3} \quad \text{for $a=1$ and as $t' \gg t_i'$}
\label{gammat}
\end{align}
\citep{gupta06,fromm13}.

Assuming no acceleration and ultrarelativistic bulk speed, then $t_i'=z_i/(\Gamma c)$, where $z_i$ is the distance from the SMBH in the jet where particle injection shuts off, and we set $t'_0=z_0/(\Gamma c)$ with $z_0=1$\,pc to facilitate comparison of $B_0'$ to core-shift derived values of the jet comoving magnetic field strength (which typically find $B_0' \approx 1$\,G at 1\,pc \citealt{pushkarev12}).  We further parametrize our model by defining the location of observed components that is furthest away from the core as $z_f$,  the initial location of observed components as $z_{\rm obs}=z_f/R_{\rm exp}$, and the ratio of the location in the jet where injection stops to where the component is first observed as $A=z_{\rm obs}/z_i$.  We introduce the parameter $A$ because the components must travel a certain distance down the jet before they can be resolved from the radio core. Now, assuming $a=1$, equation (\ref{gammat}) gives the maximum random Lorentz factor at $z_f$:
\begin{align}
\gamma_{\rm max,f} \approx 150 \left(\frac{B_0'}{\text{1\,G}}\right)^{-2}\left(\frac{z_{\rm f,pc}}{300}\right)\left(\frac{\Gamma}{10}\right)\left(\frac{R_{\rm exp}}{4}\right)^{-5/3}\left(\frac{A}{7}\right)^{-5/3}
\end{align}
with a corresponding synchrotron cutoff frequency of
\begin{align}
\nu_{\rm cut,f}\approx 12 \left(\frac{B_0'}{\text{1\,G}}\right)^{-3}\left(\frac{z_{\rm f,pc}}{300}\right)\left(\frac{\Gamma}{10}\right)^2\left(\frac{\delta}{10}\right)\left(\frac{R_{\rm exp}}{4}\right)^{-10/3}\left(\frac{A}{7}\right)^{-10/3}\text{\,GHz}
\end{align}
\citep[cf.,][]{marscher80}.  Note that the corresponding observed spectral cutoff frequency is $\nu_{\rm cut} = (3/2)\delta\gamma_{\rm max}^2eB_{\perp}'/(m_ec)$, where we assume that $B_{\perp}'\approx B'$ and that pitch angle scattering rapidly re-isotropizes the electron energy spectrum during synchrotron cooling.  The synchrotron spectrum is very well approximated by $F_{\nu}\propto \nu^{\alpha_0}\exp(-\nu/\nu_{\rm cut})$.  If we define $\alpha = d \ln {F_{\nu}}/d \ln \nu$, then
\begin{align}
\alpha = \alpha_0 - \nu/\nu_{\rm cut}.
\label{alpha}
\end{align}
Therefore, if the initial observed spectral index closer to the radio core is $\alpha(z_{\rm obs})=\alpha_0$, then we find
\begin{align}
\Delta \alpha &= \alpha(z_f) - \alpha(z_{\rm obs}) = -\frac{\nu}{\nu_{\rm cut,f}} \\
&\approx -0.53 \left(\frac{B_0'}{\text{1\,G}}\right)^{3}\left(\frac{z_{\rm f,pc}}{300}\right)^{-1}\left(\frac{\Gamma}{10}\right)^{-2}\left(\frac{\delta}{10}\right)^{-1}\left(\frac{R_{\rm exp}A}{4\times7}\right)^{10/3}\left(\frac{\nu_{obs}}{10\text{\,GHz}}\right).
\end{align}

%bibliography
\bibliographystyle{apj}
\bibliography{apj-jour,thbib}

\end{document}